\documentclass{amsart}
\usepackage[pagewise]{lineno}%\linenumbers
\usepackage[utf8]{inputenc}
\usepackage{amsthm,amsmath,amsfonts}
\usepackage{soul,color}
\usepackage{xcolor}
\usepackage{tabularx}
\usepackage{longtable}
\usepackage{booktabs}
\usepackage{mathtools}
\usepackage{mathabx}
\newcommand{\Fq}{\mathbb{F}_q}
\newtheorem{theorem}{Theorem}[section]

\newtheorem{lemma}[theorem]{Lemma}
\newtheorem{conjecture}[theorem]{Conjecture}
\newtheorem*{remark}{Remark}
\newtheorem*{ex}{Example}
\DeclareMathOperator{\ord}{Ord}
\DeclareMathOperator{\sord}{Sord}
\usepackage{color,soul,xcolor}

\usepackage{comment}
\theoremstyle{definition}
\newtheorem{definition}{Definition}[section]
\newtheorem{prop}{Proposition}[section]
\newcommand{\hlc}[2][yellow]{ {\sethlcolor{#1} \hl{#2}} }
\newcommand{\ol}[2][orange]{ {\sethlcolor{#1} \hl{#2}} }
\title{Polycyclic Codes Associated with Trinomials: Good Codes and Open Questions}
\author{Nuh Aydin, Peihan Liu, Bryan Yoshino}
\date{}

\begin{document}

\maketitle
\begin{abstract}
    Polycyclic codes are a generalization of cyclic and constacyclic codes. Even though they have been known since 1972 and received some attention more recently, there have not been many studies on polycyclic codes. This paper presents an in-depth investigation of polycyclic codes associated with trinomials. Our results include a number of facts about trinomials, some properties of polycyclic codes, and many new quantum codes derived from polycyclic codes. We also state several conjectures about polynomials and polycyclic codes. Hence, we show useful features of polycyclic codes and present some open problems related to them.
\end{abstract}

\textbf{Keywords:} polycyclic codes,  trinomials, reversible codes, quantum codes.

\section{Introduction}
A linear code $C$ of length $n$ over $\Fq$, the finite field of order $q$, is a vector subspace of $\Fq^n$. The three fundamental parameters of a linear code are: the length ($n$), the dimension ($k$) and the minimum distance ($d$). Such a code is referred to as an $[n,k,d]_q$-code. One of the most important and challenging problems in coding theory is to construct linear codes whose parameters attain the optimal values. For example, given a field $\Fq$, length $n$ and dimension $k$, constructing codes with the highest possible minimum distance $d_q[n,k]$ is one of the central problems in coding theory with many open instances. We have theoretical upper bounds on  $d_q[n,k]$ but we do not even know if it is possible to attain even the best known upper bounds. For codes over small finite fields ($q\leq 9$), there is an online database (\cite{database}) that has  information about  $d_q[n,k]$ and  best known linear codes (BKLC) along with details about their constructions. The computer algebra software Magma (\cite{magma}) also has a similar database.  

This is still a challenging problem for two main reasons: first, computing the minimum distance of an arbitrary linear code is an NP-hard problem \cite{NPhard}, and secondly the number $\displaystyle{\frac{(q^n-1)(q^n-q)\cdots (q^n-q^{k-1})}{(q^k-1)(q^k-q)\cdots (q^k-q^{k-1})}}$ of linear codes  over $\Fq$ of length $n$ and dimension $k$ grows fast as $n,k$ and $q$ increase. Therefore, it is infeasible to conduct  exhaustive searches for linear codes except for small dimensions. One way to reduce the complexity of this problem is to consider linear codes with certain algebraic structures that make them easier to analyze. A number of such classes of codes are well known including cyclic codes and their various generalizations such as   constacyclic codes, quasi-cyclic (QC) codes, and quasi-twisted (QT) codes.

 Polycyclic codes are another generalization of cyclic codes. First introduced as pseudocyclic codes in \cite{1972}, they have received some attention in the literature more recently  (see for example \cite{oto,pc2011,pc2016,pc2020}). Despite the fact that polycyclic codes have been known since 1972, they never received the same level of attention as cyclic codes and some of their generalizations. Potential reasons for this might include i) the dual of a polycyclic code is not necessarily  polycylic, ii) convenient ways of generating polycyclic codes have not been found, iii) the search space for polycyclic codes is much larger than cyclic and constacyclic codes. In \cite{seq}  there is a hint that polycyclic codes may include codes with optimal parameters. Other than a couple of examples in that work, we have not seen any examples of codes in the literature with good parameters or properties that are obtained from polycylic codes. Our goal in this work is to take a closer look at the structure and properties of polycyclic codes associated with trinomials and try to  construct linear codes with good parameters or properties from them.        
 
 Like other well known classes of codes such as constacyclic codes and QC codes, polycyclic codes are also a generalization of cyclic codes. In fact they generalize constacyclic codes too. Algebraically, a polycyclic code  is an ideal of $\Fq[x]/\langle f(x)\rangle$ for some $f(x)\in \Fq[x]$. Note that that the special case $f(x)=x^n-1$ yields cyclic codes and $f(x)=x^n-a$ gives constacyclic codes.  
 Therefore, one of our objectives in studying polycyclic codes is to expand the search space of linear codes beyond cyclic, negacyclic, and constacyclic codes. We examine the properties of self-duality, self-orthogonality, iso-duality, and reversibility for polycyclic codes that are associated  with trinomials of the form $x^n-ax^i-b$. Note that trinomials are the next step going beyond constacyclic codes.  Our investigation generated a number of pure algebra problems some of which we have been able to solve, others  stated as conjectures or open problems.

Additionally, we consider methods of constructing quantum codes from polycyclic codes  as an application.  The idea of quantum error-correcting codes was first introduced in \cite{Quantumoriginal1}, \cite{Quantumoriginal2}, and \cite{Quantumoriginal3}. Over the last couple of decades, various ways of constructing quantum codes from classical codes have been explored. The basis of most of these methods is the CSS construction that was first introduced in \cite{Quantumoriginal1} and \cite{Quantumoriginal2}. Employing the CSS construction, we have found a number of new quantum codes from polycyclic codes.

The material in this paper is organized as follows. In section two we recall some basic definitions; in section three we introduce some results about trinomials, and the duality and reversibility condition of polycyclic codes associated with trinomials. Next, we introduce quasi-polycyclic codes in section four where we generalize an important result from the quasi-twisted case.  The last two sections are about our search method for quantum error correcting codes and the new codes that we have obtained.

\begin{comment}
One of the greatest difficulties in Coding Theory is determining the minimum distance of a code. As the length and dimension of a code grows large, the computational complexity of determining minimum distance increases exponentially. One way to reduce the complexity of this problem is to construct codes with specific structures that make them easier to analyze- Cyclic, and Constacyclic Codes are some examples of linear codes which have been studied extensively. Like Constacyclic Codes are a generalization of Cyclic Codes, polycyclic Codes are a generalization of Constacyclic Codes. Therefore, the main objective of studying polycyclic Codes is to expand the search space of linear codes beyond Constacyclic Codes and to see if they can be used to construct quantum codes with good parameters.
\end{comment}
\section{Preliminaries}

\begin{definition}
  A linear code $C$ is said to be polycyclic with respect to $v = (v_0, v_1, ... , v_{n-1}) \in \Fq^n$ if for any codeword $(c_0, c_1, ..., c_{n-1}) \in C$, its right polycyclic shift, $(0,c_0,c_1,\dots, c_{n-2})+c_{n-1}(v_0,v_1,\dots,v_{n-1})$ is also a codeword. Similarly, $C$ is  \textit{left} polycyclic with respect to $v = (v_0, v_1, ... , v_{n-1}) \in \Fq^n$ if for any codeword  $(c_0, c_1, ..., c_{n-1}) \in C$, its left polycyclic shift $(c_1,c_2,\dots, c_{n-1},0)+c_0(v_0,v_1,\dots,v_{n-1})$ is also a codeword.  If $C$ is both \textit{left} and \textit{right} polycyclic, then it is \textit{bi-polycyclic}.
\end{definition}
In this work, we mainly work with right polycyclic codes, which we simply refer to as polycyclic codes. Under the usual identification of vectors with polynomials, each polycyclic code $C$ of length $n$ is associated with a vector $v$ of length $n$ (or a polynomial $v(x)$ of degree less than $n$). We call $v$ ($v(x)$) an associate vector (polynomial) of $C$, and we say that $C$ is a polycyclic code associated with $x^n-v(x)$. Moreover, polycyclic codes of length $n$  associated with $f(x)=x^n-v(x)$ are ideals of the factor ring $\Fq[x]/\langle f(x) \rangle$. Note that an associate polynomial of a polycyclic code may not be unique.

Polycyclic codes are a generalization of cyclic codes and its several important generalizations. The following are some of the most important special cases of polycyclic codes:
\begin{itemize}
\item A right polycyclic code  with respect to $v=(1,0,0,...,0)$ is a cyclic code.\\
A left polycyclic code  with respect to $v=(0,0,0...,1)$ is a cyclic code.
\item A right polycyclic code with respect to $v=(-1,0,0,...,0)$ is a  negacyclic code. A left polycyclic code with respect to  $v=(0,0,0...,-1)$ is a negacyclic code.
\item A right polycyclic code with respect to $v=(a,0,0,...,0)$ is a constacyclic code. A left polycyclic code with respect to  $v=(0,0,0,...,a^{-1})$ is a constacyclic code.
\end{itemize}

Many of the basic properties of cyclic and constacyclic codes generalize to polycyclic codes in a natural way. For example, every polycylic code $C$ over $\Fq$ of length $n$ and dimension $k$ with associated polynomial $v(x)$ has a monic polynomial $g(x)$ of minimal degree $n-k$ that belongs to $C$. This polynomial divides $x^n-v(x)$ and is called the generator polynomial of $C$ (\cite{oto}). As in the case of cyclic codes, it follows that there is a one-to-one correspondence between divisors of $x^n-v(x)$ and polycyclic codes associated with $x^n-v(x)$. We can construct a generator matrix for $C$ from its generator polynomial in exactly the same way as cyclic codes.

\begin{theorem}\cite{oto}
A code $C\subseteq \Fq^n$ is right polycyclic  with associated 
4
 polynomial $f(x)$ if and only if it has a $k \times n$ generator matrix of the form
\begin{align*}
    \begin{pmatrix}
    g_0 & g_1 &...&g_{n-k}&0&0&...&0\\
    0&g_0 &g_1&...&g_{n-k}&0&...&0\\
    \vdots&\ddots&\ddots&\ddots&\ddots&\ddots&\ddots&\ddots\\
    0&...&0&g_0&g_1&...&g_{n-k}&0\\
    0&...&0&0&g_0&g_1&...&g_{n-k}
    \end{pmatrix}
\end{align*}
with $g_{n-k}\not=0$. In this case $\langle g_0+g_1x+...+g_{n-k}x^{n-k}\rangle$ is an ideal of $\mathbb{F}[x]/\langle f(x)\rangle$.
\end{theorem}

We recall some important properties of linear codes that are of special interest in this work.
\begin{definition}\cite{mas}
    A block code $C$ is called reversible if the block of digits formed by reversing the order of the digits in a codeword is always another codeword in the same code, i,e., if $(c_0,c_1,\dots,.,c_{n-1})\in C$ then $(c_{n-1},c_{n-2},\dots, c_0)\in C$.
\end{definition}

\begin{definition} Two codes $C_1, C_2$ over $\Fq$ with generator matrices $G_1$ and $G_2$ respectively are equivalent if there exists a monomial matrix over $\Fq$ such that $MG_1=G_2$. A monomial matrix is a square matrix with exactly one non-zero entry in each row and each column.
\end{definition}

\begin{definition}
The dual of $C$, denoted by $C^\perp$, is the set of vectors orthogonal to every codeword of $C$ under the Euclidean inner product. A code $C$ is self-dual if  $C=C^{\perp}$; $C$ is iso-dual if  $C$ is equivalent to $C^{\perp}$, and $C$ is self-orthogonal if  $C\subseteq C^{\perp}$.
\end{definition}

\begin{definition}\cite{Lid}
Let $f\in\Fq[x]$ be a nonzero polynomial. If $f(0)\not=0,$ then the least positive integer $e$ for which $f(x)$ divides $x^e-1$ is called the order of $f$ and denoted by $\ord(f)=\ord(f(x)).$ If $f(0)=0$, then $f(x)=x^hg(x),$ where $h\in \mathbb{N}$ and $g\in \Fq[x]$ with $g(0)\not=0$ are uniquely determined. In this case, $\ord(f)$ is defined to be $\ord(g)$.
\end{definition}

For a prime $p$ and integers $m$ and $r$, we use the notation $p^r||m$ to denote that $r$ is the highest power of $p$ such that $p^r|m$.

\section{Trinomials}

In this paper, we  focus on right polycyclic codes associated with monic trinomials of the form $x^n - ax^i-b$ where $n>i>0$. In the sequel, whenever we refer to a trinomial we  assume the condition $n>i>0$, and unless otherwise stated by a polycyclic code we mean a polycyclic code associated with a trinomial.  In the case of cyclic codes, the factorization of $x^n-1$ over $\Fq$ gives us much information about cyclic codes of length $n$ over $\Fq$. We obtain many similar results for polycyclic codes associated with a trinomial from the factorization of the trinomial. It is well known that the polynomial $x^n-1$ does not have repeated roots over $\Fq$ if and only if $\gcd(n,q)=1$. Otherwise, every irreducible factor of $x^n-1$ over $\Fq$ has multiplicity $\gcd(n,q)$. We investigate the same question for trinomials. We will be using the fact that the Frobenius map $x\to x^{p^m}$  on a finite field $\Fq$ of characteristic $p$ is a permutation of $\Fq$ for any positive integer $m$. In particular, for any $\alpha\in \Fq$ and $m\in \mathbb{Z}^{+}$, there exists a unique $\beta\in \Fq$ such that $\alpha=\beta^{p^m}$.

\begin{comment}
\begin{lemma}
Let $\alpha,\beta\in \Fq$ and $\alpha\not=\beta$, where $q$ is a prime power. Then for all $m\in\mathbb{Z}^{+}$, $\alpha^{q^m}\not=\beta^{q^m}$.
\end{lemma}
\begin{proof}
Since $\alpha\not=\beta$, it follows that 
\begin{align*}
    (\alpha-\beta)^{q^m}&\not=0\\
    \alpha^{a^m}-\beta^{q^m}&\not=0\\
    \alpha^{a^m}&\not=\beta^{q^m}
\end{align*}
\end{proof}
\begin{lemma}
Let $q$ be a prime power and  let $\alpha\in \Fq$. Then for any $m\in \mathbb{Z}^{+}$, there exists a unique $\beta\in \Fq$ such that $\alpha=\beta^{q^m}$.
\end{lemma}
\begin{proof} Let $\Fq=\{a_1,a_2,...,a_q\}$. Given previous lemma, for any $\alpha,\beta\in \Fq$, where $\alpha\not=\beta$, we have $\alpha^{q^m}\not=\beta^{q^m}$. Hence, $\Fq=\{a_1,a_2,...,a_q\}=\{a_1^{q^m},a_2^{q^m},...,a_q^{q^m}\}$. Since $\alpha\in \Fq$, it follows that $\alpha\in\{a_1^{q^m},a_2^{q^m},...,a_q^{q^m}\}$. Therefore, there must be a unique $\beta\in \Fq$ such that $\alpha=\beta^{q^m}$.
\end{proof}
\end{comment}

\begin{lemma}
Let $g(x)=x^n-ax^i-b$ be a trinomial over $\Fq$ where $q$ and $n$ are both a power of prime $p$. Then all irreducible factors of $g(x)$ have the same multiplicity $s=\gcd(n,i)$.
\end{lemma}
\begin{proof}
Let $i=p^c\cdot d$, where $d \mod p \neq 0$ and $c,d\in \mathbb{Z}^+$, and let $n=p^m$ for some $m\in \mathbb{Z}^+$. Since $n>i$, $m>c$. Then we have
\begin{align*}
    x^n-ax^i-b&=x^{p^m}-ax^{p^c\cdot d}-b=(x^{p^{m-c}}-a'x^{d}-b')^{p^c} 
\end{align*}
%\quad \mbox{since GF($q$) is of  characteristic $p$}
where $(a')^{p^c}=a$ and where $ (b')^{p^c}=b$ by the remark before the theorem. Let's call $x^{p^{m-c}}-a'x^{d}-b'=f(x)$. Since $m>c$,  $f'(x)=p^{m-c}x^{p^{m-c}-1}-a'dx^{d-1}=-a'dx^{d-1}$. Clearly   $f(x)=x^{p^{m-c}}-a'x^{d}-b'$ is not a multiple of $x$, therefore $\gcd(f(x),f'(x))=\gcd(x^{p^{m-c}}-a'x^{d}-b',-a'dx^{d-1})$ is a constant. Thus, $f(x)=x^{p^{m-c}}-a'x^{d}-b'$ does not have any repeated factors. Hence, all irreducible factors of $f(x)$ have the same multiplicity 1. Consequently, all irreducible factors of $g(x)=x^n-ax^i-b=(f(x))^{p^{c}}$ have the same multiplicity $p^c=\gcd(n,i)=s$.
\end{proof}
We can generalize the previous result as follows.

\begin{theorem}
Let $g(x)=a_{s_0}x^{s_0}+a_{s_1}x^{s_1}+a_{s_2}x^{s_2}+...+a_{s_n}x^{s_n}\in \Fq[x]$, where $q$ is a power of prime $p$ and at most one $s_j$ is not a power of $p$, for $j=0,1,2,...,n$ $a_{s_j}\not=0$, and $a_{s_{j_1}}\not=a_{s_{j_2}}$ for $j_1,j_2=0,1,2,...,n$. Then all irreducible factors of $g(x)$ have the same multiplicity $s=\gcd(s_0,s_1,s_2,...,s_n)$.
\end{theorem}
\begin{proof}
First consider the case where exactly one  of the coefficients is not  a power of $p$. Let $s_i$ be that coefficient. So,  $s_i=p^c\cdot d$, where $d \mod p \neq 0$ and $c,d\in \mathbb{Z}$,  and for $j\not =i$, let $s_j=p^{m_j}$ for some $m_j\in \mathbb{Z}^+$. Since $char(\Fq)=p$, $g(x)$  can be written as
\begin{align*}
    g(x)&=a_{s_0}x^{s_0}+a_{s_1}x^{s_1}+a_{s_2}x^{s_2}+\cdots +a_{s_n}x^{s_n}\\
    &=a_{s_0}x^{p^{m_0}}+a_{s_1}x^{p^{m_1}}+\cdots +a_{s_i}x^{p^{c}\cdot d}+\cdots+a_{p^{m_n}}x^{s_n}\\
    &=(a_{s_0}'x^{p^{m_0-z}}+a_{s_1}'x^{p^{m_1-z}}+\cdots+a_{s_i}'x^{p^{c-z}\cdot d}+\cdots+a_{s_n}'x^{p^{m_n-z}})^{p^z},
\end{align*}
where $z=\min\{m_0,m_1,\dots,c,\dots,m_n\}$ and $(a_{s_0}')^{p^z}=a_{s_0}$. Let's call $a_{s_0}'x^{p^{m_0-z}}+a_{s_1}'x^{p^{m_1-z}}+\cdots+a_{s_i}'x^{p^{c-z}\cdot d}+\cdots+a_{s_n}'x^{p^{m_n-z}}=f(x)$. Now consider 
\begin{align*}
    \gcd(f(x),f'(x))=(a_{s_0}'x^{p^{m_0-z}}+a_{s_1}'x^{p^{m_1-z}}+...+a_{s_i}'x^{p^{c-z}\cdot d}+...+a_{s_n}'x^{p^{m_n-z}},e)=e'
\end{align*}
where $e$ and $e'$ are units. It follows that there is no repeated factors of $f(x)$, i.e., all irreducible factors of $f(x)$ have the same multiplicity 1. Therefore, all irreducible factors of $g(x)=f^{p^z}(x)$ have the same multiplicity of $p^z=s=\gcd(s_0,s_1,s_2,...,s_n)$.

If all $s_j$ are powers of $p$, then let $s_j=p^{m_j}$ for some $m_j\in \mathbb{Z}^+$. Since $char(\Fq)=p$, $g(x)$  can be written as
\begin{align*}
    g(x)&=a_{s_0}x^{s_0}+a_{s_1}x^{s_1}+a_{s_2}x^{s_2}+\cdots +a_{s_n}x^{s_n}\\
    &=a_{s_0}x^{p^{m_0}}+a_{s_1}x^{p^{m_1}}+\cdots+a_{p^{m_n}}x^{s_n}\\
    &=(a_{s_0}'x^{p^{m_0-z}}+a_{s_1}'x^{p^{m_1-z}}+\cdots+a_{s_n}'x^{p^{m_n-z}})^{p^z},
\end{align*}
where $z=\min\{m_0,m_1,\dots,m_n\}$ and $(a_{s_0}')^{p^z}=a_{s_0}$. Let's call $a_{s_0}'x^{p^{m_0-z}}+a_{s_1}'x^{p^{m_1-z}}+\cdots+a_{s_n}'x^{p^{m_n-z}}=f(x)$. Now consider
\begin{align*}
    \gcd(f(x),f'(x))=(a_{s_0}'x^{p^{m_0-z}}+a_{s_1}'x^{p^{m_1-z}}+...+a_{s_n}'x^{p^{m_n-z}},e)=e'
\end{align*}
where $e$ and $e'$ are units. It follows that there is no repeated factors of $f(x)$, i.e., all irreducible factors of $f(x)$ have the same multiplicity 1. Therefore, all irreducible factors of $g(x)=f^{p^z}(x)$ have the same multiplicity of $p^z=s=\gcd(s_0,s_1,s_2,...,s_n)$.
\end{proof}

\begin{theorem} 
For $g(x)=x^n-ax^i-b \in \Fq[x]$, where $q$ is a power of prime $p$, let $n=p^{c_1}\cdot c_2$ and $i=p^{d_1}\cdot d_2$, where $c_2\not\equiv 0 \pmod{p}$ and $d_2\not\equiv 0 \pmod{p}$. Let $z=\min\{c_1,d_1\}$,  $a=(a')^{p^z}$ and $b=(b')^{p^z}$, and let $f(x)=x^{p^{c_1-z}c_2}-a'x^{p^{d_1-z}d_2}-b'$ and $d(x)=\gcd(f(x),f'(x))$. Then 
\begin{enumerate}
    \item If $c_1\not=d_1$, then $d(x)=1$, so all irreducible factors of $g(x)=x^n-ax^i-b$ have the same multiplicity $p^z$.
    \item If $c_1=d_1$ and
    \begin{enumerate}
        \item if a $(c_2-d_2)^{th}$ root of $a'd_2c_2^{-1}$ is not a $(d_2)^{th}$ root of $b'(a')^{-1}(d_2c_2^{-1}-1)^{-1}$, then
        $d(x)=1$, so all irreducible factors of $g(x)=x^n-ax^i-b$ have the same multiplicity $s=p^z$.
        \item if not, then $d(x)\not=1$. In this case, let $D=\deg(d(x))$ and $D'=\deg(f(x))-2D=\deg(f(x))-\deg(d^2(x))$. Moreover, suppose that $d(x)$ is irreducible.
        \begin{enumerate}
            \item If $D'\not=0$, then the multiplicity of all irreducible factors of $f(x)$ that do not divide $d(x)$, is $p^z$ and $d(x)$ have multiplicity $2p^z$. Hence, the multiplicities are not the same.
            \item If $D'=0$, then all irreducible factors of $g(x)=x^n-ax^i-b$ have even multiplicity $s=2p^z$.
        \end{enumerate}
        All irreducible factors of $g(x)=x^n-ax^i-b$ either have the multiplicity $p^z$ or $2p^z$, given that $d(x)$ is an irreducible polynomial or a unit.
    \end{enumerate}
\end{enumerate}
\end{theorem}
\begin{proof}
Firstly, we are going to prove the claim (1). If $c_1\not= d_1$, then
\begin{align*}
    g(x)=x^n-ax^i-b&=x^{p^{c_1}\cdot c_2}-ax^{p^{d_1}\cdot d_2}-b\\
    &=(x^{p^{c_1-z}\cdot c_2}-a'x^{p^{d_1-z}\cdot d_2}-b')^{p^z}\\
    &=f^{p^z}(x)
\end{align*}
since $\Fq$ is of characteristic $p$. Note that either $c_1=z$ or $d_1=z$, and $c_1\not= d_1$, so $f'(x)=p^{c_1-z}c_2x^{p^{c_1-z}c_2-1}-a'p^{d_1-z} d_2x^{p^{d_1-z} d_2-1}$ is either $c_2x^{p^{c_1-z}c_2-1}$ or $-a'd_2x^{p^{d_1-z} d_2-1}$, where the only divisor of either of them is $x$. Since $f(x)=x^{p^{c_1-z}\cdot c_2}-a'x^{p^{d_1-z}\cdot d_2}-b'$ is clearly not a multiple of $x$, it follows that\\ $d(x)=\gcd(f(x),f'(x))$ is always a unit. Recall the fact that $\gcd(f(x),f'(x))=d(x)=1$ if and only if $f(x)$ has no repeated factors, so \\$f(x)=x^{p^{c_1-z}\cdot c_2}-a'x^{p^{d_1-z}\cdot d_2}-b'$ does not have any repeated factors. Hence, all irreducible factors of $f(x)$ have the same multiplicity 1, so all irreducible factors of $g(x)=x^n-ax^i-b=(f(x))^{p^z}$ have the same multiplicity $s=p^z$.

Claim (2): If $c_1=d_1$, then  $z=c_1=d_1$ and we have
\begin{align*}
    g(x)=x^n-ax^i-b&=x^{p^{c_1}\cdot c_2}-ax^{p^{d_1}\cdot d_2}-b\\
    &=(x^{c_2}-a'x^{d_2}-b')^{p^z}.
\end{align*}
Hence, $f(x)$ and $f'(x)$ are given by
\begin{align*}
    f(x)&=x^{c_2}-a'x^{d_2}-b'\\
    f'(x)&=c_2x^{c_2-1}-a'd_2x^{d_2-1}.
\end{align*}
Note that $p^{c_1}\cdot c_2=n>i=p^{d_1}\cdot d_2$ and $c_1=d_1$ implies that $c_2>d_2$, so $f'(x)$ can be written as $f'(x)=x^{d_2-1}(c_2x^{c_2-d_2}-a'd_2).$ Since $g(x)=x^n-ax^i-b=(f(x))^{p^z}$ is clearly not divisible by $x$, it follows that
\begin{align*}
    d(x)=\gcd(f(x),f'(x))&=\gcd(x^{c_2}-a'x^{d_2}-b',c_2x^{c_2-1}-a'd_2x^{d_2-1})\\
    &=\gcd(x^{c_2}-a'x^{d_2}-b',x^{d_2-1}(c_2x^{c_2-d_2}-a'd_2))\\
    &=\gcd(x^{c_2}-a'x^{d_2}-b',c_2x^{c_2-d_2}-a'd_2).
\end{align*}
Claim (2a): Let $\alpha$ be a root of $c_2x^{c_2-d_2}-a'd_2$, i.e., $\alpha$ is a $(c_2-d_2)^{th}$ root of $a'd_2c_2^{-1}$, so plugging in $\alpha$ into $x^{c_2}-a'x^{d_2}-b'$ we have the following:
\begin{align*}
    \alpha^{c_2}-a'\alpha^{d_2}-b'&=\alpha^{d_2}(\alpha^{c_2-d_2}-a')-b'\\
    &=\alpha^{d_2}(a'd_2c_2^{-1}-a')-b'\\
    &=(a'd_2c_2^{-1}-a')\cdot (\alpha^{d_2}-(a'd_2c_2^{-1}-a')^{-1}b')\\
    &=(a'd_2c_2^{-1}-a')\cdot (\alpha^{d_2}-b'(a')^{-1}(d_2c_2^{-1}-1)^{-1})
\end{align*}
Hence, $\alpha$ is a root of $x^{c_2}-a'x^{d_2}-b'$ if and only if $\alpha$ is a $(d_2)^{th}$ root of $b'(a')^{-1}(d_2c_2^{-1}-1)^{-1}$. Note that $d(x)=\gcd(f(x),f'(x))=\gcd(x^{c_2}-a'x^{d_2}-b',c_2x^{c_2-d_2}-a'd_2)=1$ if and only if $\alpha$ is not a root of $x^{c_2}-a'x^{d_2}-b'$. Thus, $d(x)=1$ if and only if a $(c_2-d_2)^{th}$ root of $a'd_2c_2^{-1}$ is not a $(d_2)^{th}$ root of $b'(a')^{-1}(d_2c_2^{-1}-1)^{-1}$. Hence, if a $(c_2-d_2)^{th}$ root of $a'd_2c_2^{-1}$ is not a $(d_2)^{th}$ root of $b'(a')^{-1}(d_2c_2^{-1}-1)^{-1}$, then $d(x)=1$. Then by the same argument all irreducible factors of $f(x)$ have the same multiplicity 1, and thus all irreducible factors of $g(x)$ have the same multiplicity $s=p^z$. 

Claim (2b): If not, then $d(x)\not=1$, so $D=\deg(d(x))>0$. Note that $d(x)|f(x)$, $d(x)|f'(x)$ and $d(x)$ being irreducible imply that $d^2(x)|f(x)$. Hence,  
 if  $D'=\deg(f(x))-2D\not=0$, then consider $f(x)/d^2(x)$. Since $d(x)$ is the greatest common divisor of $f(x)$ and $f'(x)$, there does not exist a non-constant $p(x) \mid f(x)/d^2(x)$ such that $p(x) \mid f(x)$ and $p(x) \mid f'(x)$. Hence, for all $p(x)\mid f(x)/d^2(x)$, $p^2(x)\notdivides f(x)$. In other words, all irreducible factors of $f(x)/d^2(x)$ are of multiplicity 1. Hence in this case, the multiplicity of all irreducible factors of $g(x)$ except $d(x)$ is $p^z$ and $d(x)$ have multiplicity $2p^z$. Therefore, the multiplicity of all irreducible factors of $g(x)$ definitely are not the same. If $D'=0$, then we know $f(x)=ed^2(x)$, where $e$ is a unit. Hence, all irreducible factors of $g(x)$ have the same even multiplicity of the form $2p^z$.
\end{proof}

However, over GF(2), there is a relatively simple way to determine the multiplicities of irreducible factors.

\begin{lemma} 
For $g(x)=x^m-x^i-1\in \mathbb{F}_2[x]$, all irreducible factors of $g(x)$ have the same multiplicity $2^z$, where $2^z||\gcd(n,i)$
\end{lemma}
\begin{proof}
Let $m=2^{c_1}\cdot c_2$ and $i=2^{d_1}\cdot d_2$, where $2\notdivides c_2$ and $2\notdivides d_2$.  Using a similar argument, let $z=\min\{c_1,d_1\}$  and we have
\begin{align*}
    g(x)=x^m-x^i-1&=x^{2^{c_1}\cdot c_2}-x^{2^{d_1}\cdot d_2}-1\\
    &=(x^{2^{c_1-z}\cdot c_2}-x^{2^{d_1-z}\cdot d_2}-1)^{2^z}.
\end{align*}
Let's call $x^{2^{c_1-z}\cdot c_2}-x^{2^{d_1-z}\cdot d_2}-1=f(x)$, and consider $\gcd(f(x),f'(x))$.
\begin{align*}
    \gcd(f(x),f'(x))=\gcd(x^{2^{c_1-z}\cdot c_2}-x^{2^{d_1-z}\cdot d_2}-1, 2^{c_1-z}\cdot c_2 x^{2^{c_1-z}\cdot c_2-1}-2^{d_1-z}\cdot d_2x^{2^{d_1-z}\cdot d_2-1}).
\end{align*}
Now, consider following two cases. If $c_1\not=d_1$, then only one of $c_1-z$ and $d_1-z$ is 0 and the other one is nonzero. Hence, it follows that $f'(x)=2^{c_1-z}\cdot c_2 x^{2^{c_1-z}\cdot c_2-1}-2^{d_1-z}\cdot d_2x^{2^{d_1-z}\cdot d_2-1}$ is a power of $x$. Since $f(x)$ is obviously not a multiple of $x$, we have $\gcd(f(x),f'(x))=1$.

If $c_1=d_1$, then we have 
\begin{align*}
    \gcd(f(x),f'(x))&=\gcd(x^{ c_2}-x^{ d_2}-1,  c_2 x^{ c_2-1}- d_2x^{ d_2-1})\\
    &=\gcd(x^{ c_2}-x^{d_2}-1,  x^{ c_2-1}- x^{d_2-1}) \quad\mbox{since $c_2,d_2$ are odd}.
\end{align*}
Note that $x^{ c_2-1}- x^{d_2-1}=x^{d_2-1}(x^{c_2-d_2}-1)$, so $\gcd(f(x),f'(x))=1$ if and only if for all $\alpha$ such that $\alpha^{c_2-d_2}-1=0$, $\alpha^{ c_2}-\alpha^{d_2}-1\not=0.$ Let $\alpha^{c_2-d_2}-1=0$, it follows that $\alpha^{ c_2}-\alpha^{d_2}-1=\alpha^{ d_2}(\alpha^{c_2-d_2}-1)-1 =-1\not=0$. Hence, all irreducible factors of $f(x)$ have the same multiplicity 1. Therefore, all irreducible factors of $g(x)=f^{2^z}(x)$ have the same multiplicity $2^z$, where $2^z||\gcd(n,i)$.
\end{proof}

As shown in \cite{gf7}, in certain cases there is a one-to-one correspondence between constacyclic codes associated with $x^n-a$ and those associated with $x^n-b$ such that the corresponding codes are equivalent to each other. Based on computational evidence, we conjecture analogous results for polycyclic codes associated with trinomials.
\begin{conjecture}Let $t_1(x) = x^n - ax^i - b, t_2(x) = x^n - a'x^{n-i} - b' \in \Fq[x]$ be such that $\ord(t_1(x)) = \ord(t_2(x))$. Let $C_1$ be the set of all polycyclic codes of length $n$ over $\Fq$ associated with $t_1(x)$ and $C_2$ be the set of all polycyclic codes of length $n$ over $\Fq$ associated with $t_2(x)$. Then $C_1$ and $C_2$ are in a one-to-one correspondence where corresponding codes are equivalent to each other.
\end{conjecture}
\begin{comment}
\hl{Not clear to me what the point of this next conjecture is. What is the benefit?}
\hlc[orange]{this is a weaker statement of the conjecture 3.7. I put it here because I think with this specific $\phi$ function, it is possible for us to prove it.} \hl{Is this really related to Conjecture 3.7?}
\ol{it's conjecture 3.25, the one about the code equivalence.}
The following is a weaker version of the conjecture above.\\
\begin{conjecture}
Let $f_1(x)=x^n-ax^i-b$ and $f_2=x^n-a'x^{n-i}-b'$, where $a'=a\cdot(-b)^{-1}$ and $b'=b^{-1}$. Let \hl{$\phi(x)=x^{\frac{\deg(\ord(f_1))}{\gcd(n,i)}}$}, \hlc[pink]{Are we defining $\phi$ here for the first time? also define what $\phi(f_1)$ means } then $\phi(f_1) \equiv 0 \pmod{x^{\ord(f_1)}-1}$, so \hl{$\phi(f_1) \equiv 0 \pmod{f_2}$} in the following cases:
\begin{itemize}
    \item GF3: a=b=2
    \item GF5: a=2 b=4, a=b=3, a=4 b=2
    \item GF7: a=2 b=6, a=3 b=5, a=b=4, a=5 b=3, a=6 b=2
\end{itemize}
\end{conjecture}
\end{comment}

\section{Duality}
Dual codes are of great interest in coding theory. In this section we look at the related notions of iso-dual codes, dual-containing codes, and self-orthogonal codes.  
\subsection{Iso-duality}
We  have some conjectures about iso-duality. Given these conjectures and the theorems that follow, we can produce as many iso-dual polycyclic codes as we like.

\begin{conjecture}
If a polycyclic code $C$ is generated by $g(x)|x^n-ax^i-b$ and $g^2(x)=x^n-ax^i-b$, then $C$ is isodual.
\end{conjecture}
\begin{conjecture}
Over $\mathbb{F}_2$, a polycyclic code $C$ generated by $g(x)|x^n-ax^i-b$ is iso-dual if and only if $g^2(x)=x^n-ax^i-b$. 
\end{conjecture}
\begin{theorem}[trinomial square root, characteristic not 2]
If $g(x)=x^{m}-ax^i-b\in \Fq[x]$ is a square of a polynomial $f(x)$, where $q$ is a power of prime $p$ other than 2, then $f(x)=\pm x^{m/2}\pm c$, where $c^2=-b$, $a=\pm 2c$ and $i=m/2$. 
\end{theorem}

\begin{proof}
Let $f(x)=a_{0}+a_{1}x+a_{2}x^{2}+...+a_{n}x^{n}\in \Fq[x]$. Note that $-b=a_{0}^{2}$ and $x^m=a_{n}^{2}x^{2n}$, so $a_{0}^2=-b$, $a_{n}^2=1$ and $m=2n$. Letting $a_0=c$, we have  $c^2=a_{0}^2=-b$. Clearly, $m$ has to be even and $a_{n}=\pm 1$. Now, we claim that $f(x)$ is a binomial of the form $\pm x^{m/2} \pm c$. 
\begin{align*}
    f^2(x)&=(a_{0}+a_{1}x+a_{2}x^{2}+...+a_{n}x^{n})^2\\
    &=(c+a_{s_1}x^{s_1}+a_{s_2}x^{s_2}+...\pm x^{m/2})^2 \\
    &=x^{m}-ax^i-b=g(x).
\end{align*}
Let $A(x)=a_{1}x+a_{2}x^{2}+...+a_{n-1}x^{n-1}$, then 
\begin{align*}
    f^2(x)&=(c+A(x)\pm x^{m/2})^2\\
    &=c^2+x^m+A^2(x)+2cA(x)\pm2A(x)x^{m/2}\pm 2cx^{m/2}\\
    &=-b+x^m+A^2(x)+2cA(x)\pm 2A(x)x^{m/2}\pm2cx^{m/2}\\
    &=x^{m}-ax^i-b=g(x).
\end{align*}
Thus, we have the following
\begin{align*}
   -ax^i&=A^2(x)+2cA(x)\pm 2A(x)x^{m/2}\pm 2cx^{m/2}\\
   \pm 2cx^{m/2}-ax^i&=A^2(x)+2cA(x)\pm 2A(x)x^{m/2}\\
   \pm 2cx^{m/2}-ax^i&=A(x)(A(x)+2c \pm 2x^{m/2}).
\end{align*}

Now consider the following two cases. If $\pm2cx^{m/2}-ax^i=0$ i.e., $a=\pm2c$ and $m/2=i$, then $A(x)(A(x)+2c\pm2x^{m/2})=0$, and thus either $A(x)=0$ or $A(x)+2c\pm2x^{m/2}=0$. If $A(x)=0$, then it follows that $f(x)=c+x^{m/2}$, which means we are done; if  $A(x)+2c\pm 2x^{m/2}=0$, then $A(x)=-2c\pm2x^{m/2}$, so $f(x)=-c\pm x^{m/2}$.

If $-2cx^{m/2}-ax^i\not=0$, then let $z=\min\{i,m/2\}$. It follows that $x^z\mid A(x)(A(x)+2c\pm2x^{m/2})$. Since $x\notdivides (A(x)+2c\pm2x^{m/2})$, we must have $x^z\mid A(x)$. Let $A(x)=x^z\cdot B(x)$. Then we have
\begin{align*}
    \pm2cx^{m/2}-ax^i&=A(x)(A(x)+2c\pm2x^{m/2})\\
    x^z(\pm2cx^{m/2-z}-ax^{i-z})&=x^zB(x)(x^zB(x)+2c\pm2x^{m/2})\\
    \pm2cx^{m/2-z}-ax^{i-z}&=B(x)(x^zB(x)+2c\pm2x^{m/2})
\end{align*}
Since the degree of $A(x)=x^zB(x)$ is less than $m/2$,  the degree of $B(x)(x^zB(x)+2c+2x^{m/2})$ is at least $m/2$.  Note that the degree of $-2cx^{m/2-z}-ax^{i-z}$ is either $m/2-i<m/2$ or $i-m/2<m/2$, so the highest degree of $B(x)(x^zB(x)+2c+2x^{m/2})$ is always greater than the highest degree of $-2cx^{m/2-z}-ax^{i-z}$. Thus, we reached a contradiction.

Therefore, we have shown that $f(x)$ is a binomial of the form $\pm x^{m/2}\pm c$. Hence, 
\begin{align*}
    x^m-ax^i-b=f^2(x)&=(x^{m/2}\pm c)^2\\
    &=x^m\pm2cx^{m/2}+c^2\\
    &=x^m\pm2cx^{m/2}-b
\end{align*}
and, $a=\pm 2c$.
\end{proof}

\begin{theorem}[trinomial square root, characteristic 2]
If $g(x)=x^{m}-ax^i-b\in \Fq[x]$ is the  square of a polynomial $f(x)$, where $q$ is a power of 2, then $f(x)=x^{m/2}+a'x^{i/2}+b'$, where $(a')^2=-a$, $(b')^2=-b$ and $i,m$ are even. 
\end{theorem}

\begin{proof}
Let $f(x)=a_{0}+a_{1}x+a_{2}x^{2}+...+a_{n}x^{n}\in \Fq[x]$. Note that $-b=a_{0}^{2}$ and $x^m=a_{n}^{2}x^{2n}$, so $a_{0}^2=-b$, $a_{n}^2=1$ and $m=2n$. Letting $a_0=b'$, we have  $b'^2=a_{0}^2=-b$. Obviously, $m$ has to be even and $a_{n}=\pm 1$. Hence, $f^2(x)$ is given by
\begin{align*}
    x^{m}-ax^i-b=f^2(x)&=(a_{0}+a_{1}x+a_{2}x^{2}+...+a_{n}x^{n})^2\\
    &=(b'+a_{1}x+a_{2}x^{2}+...+(\pm x^{m/2}))^2\\
    &=\sum_{j=0}^{n}(a_{j}x^{j})^2+2\sum_{\substack{j=0\\k=0\\j\not=k}}^{p}a_{j}x^{j}a_{k}x^{k}\\
    &=\sum_{j=0}^{n}(a_{j}x^{j})^2 \qquad\mbox{ since characteristic is 2}\\
    &=(a_{0})^2+(a_{n}x^{m/2})^2+\sum_{j=1}^{n-1}(a_{j}x^{j})^2\\
    &=-b+x^m+\sum_{j=1}^{p-1}(a_{s_j}x^{s_j})^2
   % &=x^{m}-ax^i-b.
\end{align*}
Then, it follows that 
\begin{align*}
    -ax^i=\sum_{j=1}^{n-1}(a_{j}x^{j})^2.
\end{align*} 
Since $\sum_{j=1}^{n-1}(a_{j}x^{j})^2$ is a sum of monomials of different degrees and there is only one term on the left side, we have $n-1=1$, so $n=2$. Therefore, $f(x)$ only has three terms i.e. $f(x)=a_0+a_1x+a_2x^2$, and thus 
\begin{align*}
    -ax^i=(a_{1}x)^2=a^{2}_{1}x^{2}.
\end{align*}
Hence, $-a=a^{2}_{1}$, which implies that $i$ is also even. Hence, $f(x)$ has the form
\begin{align*}
    f(x)=x^{m/2}+a'x^{i/2}+b',
\end{align*} where $-a=(a')^{2}$ and $-b=(b')^{2}$.

\end{proof}

Based on the theorem and the conjectures above as well as computational evidence, we have
\begin{conjecture} Over $GF(q)$,
\begin{itemize}
    \item  If $q$ is a power of an odd prime,  a polycyclic code associated with trinomial $x^{2m}\pm 2c x^m +c^2$ and generated by $g(x)=\pm x^m\pm c$  is isodual.
    \item If $q$ is a power of 2, a  polycyclic code  associated with the trinomial $x^{2m}+a^2x^{2i} +b^2$ and generated by $g(x)=x^m+ax^i+b$ is isodual.
\end{itemize}
\end{conjecture}

\begin{comment}
Based on computational evidence, we believe that a generalization of the above theorem should be true.

\begin{conjecture}
If a polynomial with  $n$ non-zero terms is a perfect square of $g(x)$, then $g(x)$ has at most $n$ terms. 
\end{conjecture}
\end{comment}

\subsection{Self-duality, Dual-containing and Self-orthogonality}

Self-dual, self-orthogonal and dual-containing linear codes are widely used to construct quantum codes. For example, CSS construction requires
two linear codes $C_1$ and $C_2$ such that $C_2^{\perp}\subseteq C_1$. Hence, if $C$ is self-dual, then we can construct a CSS quantum code using $C$ alone since $C^{\perp}\subseteq C$. Similarly, if  $C$ is self-orthogonal, then we can construct a CSS quantum code using  $C_1=C^{\perp}$ and $C_2=C$   since $C_2^{\perp}=C\subseteq C_1=C^{\perp}$. We have a similar situation if $C$ is dual-containing.

For self-duality and self-orthogonality, we came up with two conjectures.  Based on the following theorems and assuming the conjectures, we can characterize all self-dual polycyclic codes that are actually constacyclic codes. It is worth recalling that the dual of a polycyclic code is not necessarily polycyclic, however, it is always a sequential code \cite{oto,seq}. It is also shown in \cite{oto} that constacyclic codes are those that are both polycyclic and sequential.
\begin{lemma}
Let $g(x)=a_0+a_1x+\cdots +a_sx^s|x^n-ax^i-b$ be the generator polynomial of a polycyclic code $C$ of length $n$, and let $e=\ord(x^n-ax^i-b)$ and $h(x)=b_0+b_1x+...+b_{e-s}x^{e-s}=\frac{x^e-1}{g(x)}$. Then a parity check matrix of $C$ is given by
\begin{align*}
    \begin{pmatrix}
    b_{e-s} & b_{e-s-1} &b_{e-s-2}&...&...&...&... \\
    0&b_{e-s} & b_{e-s-1} &b_{e-s-2}&...&...&...\\
    \vdots&\ddots&\ddots&\ddots &\ddots&\ddots&\cdots\\
    0&...&0&b_{e-s} & b_{e-s-1} &b_{e-s-2}&...&\\
    \end{pmatrix}_{s\times n}
\end{align*}
which is a submatrix of the parity check matrix of the cyclic code of length $e$ generated by $g(x)=a_0+a_1x+...+a_sx^s|x^e-1$.
\end{lemma}
\begin{proof}
Let the generator polynomial of the polycyclic code be of degree $s$   and the length of this code  be $n$, then a generator matrix is given by
\begin{align*}
G=
\begin{pmatrix}
a_0 &a_1 &a_2&...&a_{s-1}&0&0&...\\
0 &a_0 &a_1&...&a_{s-2}&a_{s-1}&0&...\\
\vdots &\ddots & \ddots & \ddots &\ddots&\ddots&\ddots &\ddots&\\
0&...&0&a_0&a_1&...&a_{s-2}&a_{s-1}
\end{pmatrix}_{n-s \times n}.
\end{align*}
The reversed code $C'$ has a generator matrix
\begin{align*}
G'=
\begin{pmatrix}
...&0&0&a_{s-1}&...&a_2&a_1&a_0\\
...&0&a_{s-1}&a_{s-2}&...&a_1&a_0&0\\
 \vdots & \vdots &\vdots&\vdots&\vdots &\vdots&\vdots&\vdots\\
a_{s-1}&a_{s-2}&...&a_1&a_0&0&...&0
\end{pmatrix}_{n-s \times n}.
\end{align*}
If each row in $G$ is orthogonal to every other rows, then clearly each row in $G'$ is orthogonal to every other rows since $\mbox{row}_{G_i}\cdot\mbox{row}_{G_j}=\mbox{row}_{G'_i}\cdot\mbox{row}_{G'_j}$. Hence, if $C$ is self-orthogonal, then $C'$ is also self-orthogonal.

Furthermore, let the generator matrix of the polycyclic code generated by $g^*$ be $G''$. It is given by
\begin{align*}
G''=
\begin{pmatrix}
a_{s-1} &a_{s-2} &a_{s-3}&...&a_0&0&0&...\\
0 &a_{s-1} &a_{s-2}&...&a_{1}&a_{0}&0&...\\
\vdots &\ddots & \ddots & \ddots &\ddots&\ddots&\ddots &\vdots\\
0&...&0&a_{s-1}&a_{s-2}&...&a_{1}&a_{0}
\end{pmatrix}_{n-s \times n}.
\end{align*}
Hence, by a  permutation of the rows, it is easy to see that $G'=G''$, so $C'$ is generated by $g^*$.
\end{proof}

Based on computational evidence, we have another conjecture about self-dual polycyclic codes.

\begin{conjecture}\label{:sdconj}
A polycyclic code over $\Fq$ is self-dual if and only if its generator is in the form $g=x^m-a|x^{2m}-bx^m-c$, where $a^2=q-1$. Also, the minimum distance of all self-dual polycyclic codes is 2 and they are all actually constacyclic codes.
\end{conjecture}
\begin{comment}
\begin{conjecture}
A polycyclic code over $\Fq$ is self-dual constacyclic code if and only if its generator is in the form  $g=x^m-a|x^{2m}-bx^m-c$, where $a^2=q-1$.
\end{conjecture}
\end{comment}

\begin{lemma}
Suppose the above conjecture is true. Then self-dual polycyclic codes only exist over GF($2^n$) and GF($p^n$) (except for GF(2)), where $p\equiv 1 \pmod{4}$. 
\end{lemma}
\begin{proof}
Given the conjecture \ref{:sdconj}, a self-dual polycyclic code exists if $a^2=-1 \pmod{q}$ has a solution. For characteristic $p=2$, obviously $a=1$ is always a solution. However, over $GF(2)$ there does not exists   $g=x^m-1|x^{2m}-x^m-1$ because $x^{2m}-x^m-1 \equiv 1 \pmod{x^m-1}$. So self-dual polycylic codes only exist in GF($2^n$), where $n>1$ for $a=1$.  For any other $q$ that is a power of an odd prime $p$, we have $a^2\equiv -1 \pmod{p^n}$ has a solution if and only if $a^2\equiv -1 \pmod{p}$ has solution. Hence, by the definition of Legendre symbol, we know $a^2=-1 \pmod{q}$ having a solution is equivalent to $(\frac{-1}{p})=1$. Note that $(\frac{-1}{p})=1$ if and only if $p\equiv 1 \pmod{4}$. Therefore, self-dual polycyclic codes only exist in $GF(q=p^n)$, where $p\equiv 1 \pmod{4}$.
\end{proof}
\begin{remark}
For small fields whose size are less than or equal to 19, self-dual polycyclic codes only exist over $GF(4), GF(5), GF(8), GF(13), GF(16), GF(17)$, and the constant term of the generator polynomials is either 2 or 3 for $GF(5^n)$, 5 or 8 for $GF(13)$, 4 or 13 for $GF(17)$ and 1 for $GF(2^n)$.
\end{remark}

The following result can easily be proven much like cyclic codes.
\begin{theorem}
If a polycyclic code $C$ is self-orthogonal, then its reversed code $C'$ is also self-orthogonal.  Also its reversed code is generated by the reciprocal polynomial of the generator polynomial of $C$. Furthermore, $C'$ is always equivalent to $C$ because of the isomorphism.
\end{theorem}
\begin{comment}
\begin{proof}
Let a generator matrix of  polycyclic code $C$  be
\begin{align*}
G=
\begin{pmatrix}
a_0 &a_1 &a_2&...&a_{s-1}&0&0&...\\
0 &a_0 &a_1&...&a_{s-2}&a_{s-1}&0&...\\
\vdots &\ddots & \ddots & \ddots &\ddots&\ddots&\ddots &\ddots&\\
0&...&0&a_0&a_1&...&a_{s-2}&a_{s-1}
\end{pmatrix}_{n-s \times n}.
\end{align*}
Then the reversed code $C'$ has a generator matrix
\begin{align*}
G'=
\begin{pmatrix}
...&0&0&a_{s-1}&...&a_2&a_1&a_0\\
...&0&a_{s-1}&a_{s-2}&...&a_1&a_0&0\\
 \vdots & \vdots &\vdots&\vdots&\vdots &\vdots&\vdots&\vdots\\
a_{s-1}&a_{s-2}&...&a_1&a_0&0&...&0
\end{pmatrix}_{n-s \times n}.
\end{align*}
If each row in $G$ is orthogonal to all other rows, then clearly each row in $G'$ is orthogonal to all other rows as well since $\mbox{row}_{G_i}\cdot\mbox{row}_{G_j}=\mbox{row}_{G'_i}\cdot\mbox{row}_{G'_j}$. Hence, if $C$ is self-orthogonal, then $C'$ so is.

Now consider the polycyclic code generated by $g^*$. A generator generator matrix of this code  is 
\begin{align*}
G''=
\begin{pmatrix}
a_{s-1} &a_{s-2} &a_{s-3}&...&a_0&0&0&...\\
0 &a_{s-1} &a_{s-2}&...&a_{1}&a_{0}&0&...\\
\vdots &\ddots & \ddots & \ddots &\ddots&\ddots&\ddots &\vdots\\
0&...&0&a_{s-1}&a_{s-2}&...&a_{1}&a_{0}
\end{pmatrix}_{n-s \times n}.
\end{align*}
Hence, by rows permutation, it is easy to see that $G'=G''$, so $C'$ is generated by $g^*$.
\end{proof}

\end{comment}

\begin{conjecture}
There is no self-dual or self-orthogonal or dual-containing polycyclic codes over the binary field.
\end{conjecture}
\begin{conjecture}
A polycyclic code is not dual-containing if the generator polynomial $g|x^n-ax^i-b$, where $n$ is prime.
\end{conjecture}

Here are a few examples of self-dual polycyclic codes that we obtained.
\begin{longtable}{p{3cm}p{.5cm}p{.5cm}p{.5cm}p{.5cm}p{4.5cm} }

\caption{polycyclic Codes that are Self-Dual}
\\\hline\noalign{\smallskip}
$[n,k,d]_{q}$  &$n$& $i$ & $a$ & $b$ & Polynomial  \\
\noalign{\smallskip}\hline\noalign{\smallskip}
   $[38, 19, 2]_{4}$ &38 & 19 & $\alpha^2$ &$\alpha$ &[10000000000000000001] \\
    $[18, 9, 2]_{5}$ &18 & 9 & 1 &2 & [1000000003] \\
       $[20, 10, 2]_{5}$ & 20 & 10 & 1 &1 & [10000000002] \\
    $[22, 11, 2]_{13}$ &22 & 11 & 9 &6 & [100000000008] \\
\noalign{\smallskip}\hline
\end{longtable}

\bigskip

\section{Reversibility}
Reversible codes have essential applications for DNA computing \cite{dna2}. Each single DNA strand is composed with a sequence of nucleotides, and it can be paired up with a complementary strand to form a double helix\cite{dna}. Finding reversible codes is one of the most essential requirements for codes suitable for DNA computing. Also, in certain data storage applications, the reversible transformation is relatively easy.

We came up some useful theorems to find the reversible polycyclic codes and the number of reversible polycyclic codes from a given trinomial. First, we introduce some relevant results about self-reciprocal and semi-reciprocal polynomials.

%\hl{Must define semi reciprocal polynomial used in the next theorem. You do it later but it is too late.}

\begin{definition}
For any $f(x)=a_0+a_1x+\cdots a_{n-1}x^{n-1}+a_nx^n \in \Fq[x]$ its reciprocal polynomial $f^{*}(x)$  is defined by 
$$ f^{*}(x)=x^nf({1 \over x})=a_n+a_{n-1}x+\cdots+a_1x^{n-1}+a_0x^n.$$
If $f(x)^{*}=f(x)$, then $f(x)$ is called self-reciprocal.
\end{definition}

 It is well known that (\cite{mas}) a cyclic code $C=\langle g(x)\rangle$ is reversible if and only if $g(x)$ is self-reciprocal. Now we generalize this notion which will be needed for polycyclic codes.

\begin{definition}
A polynomial $f(x)$ is semi-reciprocal if  $f(x)=\alpha f^*(x)$, where $\alpha$ is a constant other than 1 and $f^*(x)$ is the reciprocal polynomial of $f(x)$.
\end{definition}

\begin{prop}
If $f_i$ for $i=1,2,...,n$ are self-reciprocal, then $\prod f_i$ is also self-reciprocal.
\end{prop}
\begin{proof}
Consider $(\prod f_i)^*$. Since $(\prod f_i)^*=\prod f_i^*=\prod f_i$,  $\prod f_i$ is self-reciprocal.
\end{proof}

\begin{prop}
For any polynomial $f(x)$, $\deg(f(x))\not= \deg(f^*(x))$ if and only if $x|f(x)$. Also, $\deg(f(x))-\deg(f^*(x))=k$, where $k$ is the greatest integer such that $x^k|f(x)$.
\end{prop}
\begin{proof}
Suppose $x|f(x)$, then we can write $f(x)$ as $f(x)=x^k\cdot g(x)$, where $x\notdivides g(x)$, i.e., $g(x)$ has a non-zero constant term $c$. The reciprocal polynomial of $f(x)$ is given by
\begin{align*}
    f^*(x)=x^{\deg(f(x))}f(\frac{1}{x})&=x^{\deg(f(x))}\cdot(\frac{1}{x^k}\cdot g(\frac{1}{x}))\\
    &=x^{\deg(f(x))-k}\cdot g(\frac{1}{x})
\end{align*}
Note that the  degree of $g(\frac{1}{x})$ is 0, so the degree of $f^*(x)=x^{\deg(f(x))-k}\cdot g(\frac{1}{x})$ is  $\deg(f(x))-k$. 

Suppose $\deg(f(x))\not= \deg(f^*(x))$. Note that the degree of $f^*=x^{\deg(f)}f(\frac{1}{x})$ is less than $\deg(f)$, so let $k=\deg(f(x))-\deg(f^*(x))$. Let $f(x)=a_0+a_1x+...+a_nx^n$, where $a_n\not=0$. It follows that
\begin{align*}
    f^*(x)&=x^{n}\cdot f(\frac{1}{x})=a_n+a_{n-1}x+...+a_1x^{n-1}+a_0x^n.
\end{align*}
Since the degree of $f^*$ is $\deg(f(x))-k$, we must have $a_0=a_1=...=a_{k-1}=0$, and thus $f(x)$ is given by 
\begin{align*}
    f(x)&=a_0+a_1x+...+a_nx^n\\
    &=a_kx^k+a_{k+1}x^{k+1}+...+a_nx^n\\
    &=x^k(a_k+a_{k+1}x^{1}+...+a_nx^{n-k}).
\end{align*}
Therefore, $x^k|f(x)$.
\end{proof}

\begin{prop}
If $f(x)$ is semi-reciprocal, then $f^2(x)$ is self-reciprocal if and only if $\alpha^2=1$.
\end{prop}
\begin{proof}
Consider $$(f^2(x))^*=f^*(x)\cdot f^*(x)=\alpha^{-1}f(x) \cdot \alpha^{-1} f(x)=\alpha^{-2}f^2(x) $$

Hence, $f^2(x)$ is self-reciprocal if and only if $\alpha^{-2}=1$ if and only if $\alpha^2=1$.
\end{proof}
\begin{prop}
If $f(x)$ is semi-reciprocal, where $f^*(x)\cdot\alpha=f(x)$, then $\alpha^2=1$.
\end{prop}
\begin{proof}
Since $f(x)=\alpha f^*(x)$, we have
\begin{align*}
    f^*(x)&=(\alpha f^*(x))^*=\alpha f(x).
\end{align*}
Hence, given that $f(x)=\alpha f^*(x)$ and $f^*(x)=\alpha f(x)$, it follows that $\alpha^2=1$.
\end{proof}

Combining the two propositions above, we have
\begin{prop} If $f(x)$ is semi-reciprocal, then $f^2(x)$ is self-reciprocal.
\end{prop}

\begin{proof}
Since the square root of unity is either 1 or -1, by definition we must have $\alpha=-1$. Given two semi-reciprocal polynomials $f(x)$ and $g(x)$  where $-f^*(x)=f(x)$ and $-g^*(x)=g(x)$, we have $g(x)f(x)=g^*(x)f^*(x)=(g(x)f(x))^*.$ Taking $g(x)=f(x)$, the result follows.
\end{proof}

\begin{remark}
If $f(x)$ is semi-reciprocal, where $f^*(x)\cdot\alpha=f(x)$, then $\alpha=-1$. Therefore, the product of any two semi-reciprocal polynomials is self-reciprocal.
\end{remark}

The next theorem shows that the generalization of the well known result about reversibility of cyclic codes involves the notion of semi reciprocal polynomial.

\begin{theorem}
A polycyclic code generated by $g(x)|x^{n}-ax^{i}-b$ is reversible if and only if $g(x)$ is a semi reciprocal polynomial.
\end{theorem}
\begin{proof}
Let $C$ be the polycylic code generated by $g(x)=g_0+g_1x+\cdots +g_rx^r\mid x^n-ax^i-b$ of degree $r$. Then  $ C=\{ (f_0,f_1,...,f_{n-1}): g(x)\mid f(x)\}$ and the dimension of $C$ is $n-r$. Let $f(x)=g(x)j(x)$ be an arbitrary codeword where $\deg(j(x))\leq n-r-1$.

Now consider the reversed codeword, ($f_{n-1},...,f_1,f_0$) which is given by $f_R(x)=x^{n-1}f(x^{-1})$. Since $f(x)=g(x)j(x)$, it follows that
\begin{align*}
    f_R(x)&=x^{n-1}f(x^{-1})\\
    &=x^{n-1}\cdot g(x^{-1})j(x^{-1})\\
    &=x^{n-r-1}j(x^{-1}) \cdot x^rg(x^{-1})\\
    &=\alpha^{-1} x^{n-r-1}j(x^{-1}) \cdot \alpha x^rg(x^{-1}).
\end{align*}
Note that $x^rg(x^{-1})$ is the reciprocal polynomial of $g(x)$ and $\alpha x^rg(x^{-1})$ is a semi-reciprocal polynomial of $g(x)$, denoted by $g'^*(x)$. Hence, an arbitrary codeword in the reversed code is of the form $p(x)\cdot (g_0^{-1}x^rg(x^{-1}))=p(x)\cdot g'^*(x)$ for some polynomial $p(x)$ of degree $\leq n-r-1$. Therefore the polycyclic code is reversible if and only if $g(x)=g'^*(x)$, i.e., $g(x)$ is semi-reciprocal.
\end{proof}

\begin{definition}
Two polynomials $f(x)$ and $g(x)$ are pair-reciprocal (or mutually reciprocal) if $f(x)=\alpha g^*(x)$, where $\alpha$ is a constant and $g^*(x)$ is the reciprocal polynomial of $g(x)$.
\end{definition}
\begin{prop}
If $f(x)$ and $g(x)$ are pair-reciprocal, then $f(x)g(x)$ is self-reciprocal.
\end{prop}
\begin{proof}
Let $f(x)=\alpha g^*(x)$, where $\alpha$ is a constant and $g^*(x)$ is the reciprocal polynomial of $g(x)$. Then we have $\displaystyle{    f^*(x)=(\alpha g^*(x))^*=\alpha g(x).}$ Given that $f(x)=\alpha g^*(x)$ and $f^* (x)=\alpha g(x)$, we have
\begin{align*}
   f^*(x)g^*(x)=\alpha g(x) \cdot {1\over \alpha}f(x)= f(x)g(x)
\end{align*}
Hence, $f(x)g(x)$ is self-reciprocal.
\end{proof}
\begin{remark}
If $f$ is semi reciprocal or self reciprocal, then $x\notdivides f$. If $f$ and $g$ are pair reciprocal, then $x\notdivides f$ and $x\notdivides g$.
\end{remark}
\begin{proof}
If $f$ is semi reciprocal or self reciprocal, then $\deg(f)=\deg(f^*)$, so $x\notdivides f$. Similarly, if $f$ and $g$ are pair reciprocal, then $\deg(f)=\deg(g)$, so $x\notdivides f$ and $x\notdivides g$.
\end{proof}

\begin{lemma}
Let $g_1,g_2$ be two polynomials. Let $$g_   1=(p_1^{s_1}p_2^{s_2}...p_A^{s_A})\cdot (q_1^{m_1}\Bar{q}_{1}^{m_1'} q_2^{m_2}\Bar{q}_{2}^{m_2'}... q_B^{m_B}\Bar{q}_{B}^{m_B'})\cdot (f_1^{w_1}f_2^{w_2}...f_C^{w_C})\cdot D$$ where each $p_i$ is a self reciprocal irreducible polynomial, $q_i$'s and $\Bar{q_i}$'s are pair-reciprocal irreducible polynomials,  each $f_i$ is a semi reciprocal irreducible polynomial, and $D$ is a polynomial with none of the features above. Let $n_i=\min\{m_i,m_i'\}$ and $2e_i=\lfloor\frac{w_i}{2}\rfloor$. Then 
\begin{enumerate}
    \item $g_1g_2$ is self reciprocal if (and only if)
    \begin{align*}
        g_2=D^*(q_1^{m_1'-n_1}\Bar{q}_{1}^{m_1-n_1} q_2^{m_2'-n_2}\Bar{q}_{2}^{m_2-n_2}... q_B^{m_B'-n_B}\Bar{q}_{B}^{m_B-n_B})\\
    \cdot (f_1^{w_1-2e_1}f_2^{w_2-2e_2}...f_C^{w_C-2e_C})\cdot E,
    \end{align*}
    where $E$ is a self reciprocal polynomial.
    
    \item $g_1g_2$ is semi reciprocal if (and only if)
    \begin{align*}
        g_2=\alpha D^*(q_1^{m_1'-n_1}\Bar{q}_{1}^{m_1-n_1} q_2^{m_2'-n_2}\Bar{q}_{2}^{m_2-n_2}... q_B^{m_B'-n_B}\Bar{q}_{B}^{m_B-n_B})
    \cdot E,
    \end{align*}
    where $E$ is a semi reciprocal polynomial and $\alpha$ is a constant.
\end{enumerate}
\end{lemma}
\begin{proof}
Assume 
\begin{align*}
    g_2=D^*(q_1^{m_1-n_1}\Bar{q}_{1}^{m_1'-n_1} q_2^{m_2-n_2}\Bar{q}_{2}^{m_2'-n_2}... q_B^{m_B-n_B}\Bar{q}_{B}^{m_B'-n_B})\\
\cdot (f_1^{w_1-2e_1}f_2^{w_2-2e_2}...f_C^{w_C-2e_C})\cdot E,
\end{align*}
where $E$ is a self reciprocal polynomial. Then $g_1g_2$ is given by
\begin{align*}
    g_1g_2&=(p_1^{s_1}p_2^{s_2}...p_A^{s_A})\cdot (q_1^{m_1}\Bar{q}_{1}^{m_1'} q_2^{m_2}\Bar{q}_{2}^{m_2'}... q_B^{m_B}\Bar{q}_{B}^{m_B'})\cdot (f_1^{w_1}f_2^{w_2}...f_C^{w_C})\cdot D\\
    &\cdot D^*(q_1^{m_1'-n_1}\Bar{q}_{1}^{m_1-n_1} q_2^{m_2'-n_2}\Bar{q}_{2}^{m_2-n_2}... q_B^{m_B'-n_B}\Bar{q}_{B}^{m_B-n_B})\\
    &\cdot (f_1^{w_1-2e_1}f_2^{w_2-2e_2}...f_C^{w_C-2e_C})\cdot E\\
    &=(DD^*)\cdot (E\cdot p_1^{s_1}p_2^{s_2}...p_A^{s_A})\\ &\cdot (q_1^{m_1'+m_1-n_1}\Bar{q}_{1}^{m_1+m1'-n_1} q_2^{m_2'+m_2-n_2}\Bar{q}_{2}^{m_2+m_2'-n_2}... q_B^{m_B'+m_B-n_B}\Bar{q}_{B}^{m_B+m_B'-n_B})\\
    &\cdot (f_1^{2w_1-2e_1}f_2^{2w_2-2e_2}...f_C^{2w_C-2e_C})
\end{align*}
Note that $DD^*$ and\\ $E\cdot p_1^{s_1}p_2^{s_2}...p_A^{s_A}$ are self-reciprocal, and\\
$q_1^{m_1'+m_1-n_1}\Bar{q}_{1}^{m_1+m1'-n_1} q_2^{m_2'+m_2-n_2}\Bar{q}_{2}^{m_2+m_2'-n_2}... q_B^{m_B'+m_B-n_B}\Bar{q}_{B}^{m_B+m_B'-n_B}$ and\\ $(f_1^{2w_1-2e_1}f_2^{2w_2-2e_2}...f_C^{2w_C-2e_C})$ are self-reciprocal by propositions above. It follows that $g_1g_2$ is self-reciprocal.

Now for semi-reciprocal condition, we assume 
\begin{align*}
    g_2=\alpha D^*(q_1^{m_1'-n_1}\Bar{q}_{1}^{m_1-n_1} q_2^{m_2'-n_2}\Bar{q}_{2}^{m_2-n_2}... q_B^{m_B'-n_B}\Bar{q}_{B}^{m_B-n_B})
    \cdot E,
\end{align*}
where $E$ is a semi reciprocal polynomial and $\alpha$ is a constant. Then $g_1g_2$ is given by
\begin{align*}
    g_1g_2&=(p_1^{s_1}p_2^{s_2}...p_A^{s_A})\cdot (q_1^{m_1}\Bar{q}_{1}^{m_1'} q_2^{m_2}\Bar{q}_{2}^{m_2'}... q_B^{m_B}\Bar{q}_{B}^{m_B'})\cdot (f_1^{w_1}f_2^{w_2}...f_C^{w_C})\cdot D\\
    &\alpha D^*(q_1^{m_1'-n_1}\Bar{q}_{1}^{m_1-n_1} q_2^{m_2'-n_2}\Bar{q}_{2}^{m_2-n_2}... q_B^{m_B'-n_B}\Bar{q}_{B}^{m_B-n_B})
    \cdot E\\
    &= \alpha(DD^*\cdot (p_1^{s_1}p_2^{s_2}...p_A^{s_A}))\\
    &\cdot ((q_1^{m_1'+m_1-n_1}\Bar{q}_{1}^{m_1+m1'-n_1} q_2^{m_2'+m_2-n_2}\Bar{q}_{2}^{m_2+m_2'-n_2}... q_B^{m_B'+m_B-n_B}\Bar{q}_{B}^{m_B+m_B'-n_B}))\\
    &\cdot (E(f_1^{w_1}f_2^{w_2}...f_C^{w_C}))
\end{align*}

Note that  $((q_1^{m_1'+m_1-n_1}\Bar{q}_{1}^{m_1+m1'-n_1} q_2^{m_2'+m_2-n_2}\Bar{q}_{2}^{m_2+m_2'-n_2}... q_B^{m_B'+m_B-n_B}\Bar{q}_{B}^{m_B+m_B'-n_B}))$ and $DD^*\cdot (p_1^{s_1}p_2^{s_2}...p_A^{s_A})$ are self reciprocal, and $\alpha E(f_1^{w_1}f_2^{w_2}...f_C^{w_C})$ is self-reciprocal. Hence, $g_1g_2$ is semi reciprocal.
\end{proof}

\begin{theorem}
Given a trinomial $x^n-ax^i-b$, we can write it as 
\begin{align*}
    x^n-ax^i-b=&p_1^{s_1}\cdot p_2^{s_2}\cdot ... \cdot p_A ^{s_A}\\
    &\cdot q_1^{m_1}\Bar{q}_{1}^{m_1'}\cdot \cdot q_2^{m_2}\Bar{q}_{2}^{m_2'}\cdot...\cdot q_B^{m_B}\Bar{q}_{B}^{m_B'}\\
    &\cdot f_1^{w_1}\cdot f_2^{w_2}\cdot... \cdot f_C^{w_C}\\
    &\cdot D
\end{align*}
where $p_i$'s are self-reciprocal irreducible factors, $q_i$'s and $\Bar{q_i}$'s are pair-reciprocal polynomials, $f_i$'s are semi-reciprocal polynomials, and $D$ is a polynomial with none of the features above. Let $n_i=\min\{m_i,m_i'\}$ and $2e_i=\lfloor\frac{w_i}{2}\rfloor$. Then the number of self-reciprocal factors of the trinomial is
\begin{align*}
    \prod (s_i+1) \cdot \prod (n_i+1) \cdot \prod (e_i+1) 
\end{align*}
\end{theorem}
\begin{proof}
The proof  directly follows from the preceding  lemmas.
\end{proof}
\begin{remark}
Given a trinomial $x^n-ax^i-b$ or a binomial $x^n-a$, and we can write it in the form
\begin{align*}
    x^n-ax^i-b=&p_1^{s_1}\cdot p_2^{s_2}\cdot ... \cdot p_A ^{s_A}\\
    &\cdot q_1^{m_1}\Bar{q}_{1}^{m_1'}\cdot \cdot q_2^{m_2}\Bar{q}_{2}^{m_2'}\cdot...\cdot q_B^{m_B}\Bar{q}_{B}^{m_B'}\\
    &\cdot f_1^{w_1}\cdot f_2^{w_2}\cdot... \cdot f_C^{w_C}\\
    &\cdot D.
\end{align*} Then, there are $\prod (s_i+1) \cdot \prod (n_i+1) \cdot \prod (e_i+1)$ reversible polycyclic/constacyclic codes (including trivial generator 1)
\end{remark}

\begin{remark}
For $f_1(x)=x^n-ax^i-b$ and $f_2(x)=x^n-a'x^i-b'$, where $a'=a(-b)^{-1}$ and $b=b^{-1}$, we have $\gcd(f_!(x),f_2(x))=\prod p_i^{s_i}(x)$, where $p_i(x)$'s are the semi-reciprocal factors of $f_1(x)$
\end{remark}

\section{On the Order of Trinomials}

\begin{theorem}[same order]
Let trinomials $f_1=x^n-ax^i-b$ and $f_2=x^n-a'x^{n-i}-b'$, where $a'=a\cdot(-b)^{-1}$ and $b'=b^{-1}$, be mutually reciprocal.   Then $\ord(f_1)=\ord(f_2).$
\end{theorem}
\begin{proof}
The reciprocal polynomial of $f_1$ is given by $f_1^*=-bx^n-ax^{n-1}+1$. Consider $f_1'^{*}=f_1^* \cdot (-b)^{-1}=x^n-a(-b)^{-1}x^{n-1}+(-b)^{-1}=f_2$. Note that $\ord(f_1)=\ord(f_1^*)$ and $\ord(f_1^*)=\ord(f_1'^*)=\ord(f_2)$, so it follows that $\ord(f_1)=\ord(f_2).$ Also note that $f_1$ and $f_2$ are actually pair reciprocal. Hence, this proof also tells us that if $f_1$ and $f_2$ are pair reciprocal, then $\ord(f_1)=\ord(f_2).$
\end{proof}

\begin{theorem}[same degree distribution]
If $f_1(x)$ and $f_2(x)$ are mutually reciprocal, then the degree distributions of the irreducible factors of $f_1(x)$ and $f_2(x)$ are the same. In particular, for trinomials  $f_1(x)=x^n-ax^i-b$ and $f_2=x^n-a'x^{n-i}-b'$, where $a'=a\cdot(-b)^{-1}$ and $b'=b^{-1}$,  the degree distributions of irreducible factors of $f_1(x)=x^n-ax^i-b$ and $f_2(x)=x^n-a'x^{n-i}-b'$ are the same.
\end{theorem}
\begin{proof}
Let $f_1(x)=\prod_{i=1}^{s}p_i^{r_i}(x)$, where $\forall i$, $p_i(x)$ is irreducible, and let $f_2=\alpha f_1^*.$ Consider $f_1^*(x)$:
\begin{align*}
    f_1^*(x)&=x^{\deg(f_1(x))}f_1(\frac{1}{x})\\
    &=x^{\deg(f_1(x))}\prod_{i=1}^{s}p_i^{r_i}(\frac{1}{x})\\
    &=\prod_{i=1}^{s}x^{\deg(p_i^{r_i}(x))}p_i^{r_i}(\frac{1}{x})\\
    &=\prod_{i=1}^{s}x^{r_i\cdot \deg(p_i(x))}p_i^{r_i}(\frac{1}{x})\\
    &=\prod_{i=1}^{s}\left(x^{\deg(p_i(x))}p_i(\frac{1}{x})\right)^{r_i}\\
    &=\prod_{i=1}^{s}\left(p_i^*(x)\right)^{r_i}
\end{align*}
Hence, $f_2=\alpha \prod_{i=1}^{s}(p_i^*(x))^{r_i}$. Note that $x\notdivides f_1$ since $f_1$ and $f_2$ are pair reciprocal, so $x\notdivides p_i$ for all $p_i$. Hence, $\deg(p_i)=\deg(p_i^*)$. Therefore, the degrees of irreducible factors of $f_2$ are the same as the degrees of irreducible factors of $f_1$, and the multiplicities are also the same. Thus, the degree distributions of $f_1$ and $f_2$ are the same.
\end{proof}

\begin{theorem}
For any polynomial $f=\sum_{a_j\not=0}a_jx^{s_j}$, $\gcd(s_j)|\ord(f)$
\end{theorem}
\begin{proof}
Let $f(x)=\sum_{a_j\not=0}a_jx^{s_j}$ and $d=\gcd(s_j)$. Then we can rewrite it as $f(x)=b_0+b_1x^d+b_2x^{2d}+...+b_nx^{nd}$, so $f(x^d)$ can be written as $f(x^d)=b_0+b_1x+b_2x^{2}+...+b_nx^{n}$. Let $e$ be the order of $f(x)$, and $h(x)=\frac{x^e-1}{f(x)}=c_0+c_1x+x_2x^2+...+c_mx^m$, and let $C$ be the cyclic code generated by $f(x)$. 
Let $e'$ be the order of $f(x^d)$.
Then a generator matrix of the cyclic code of length $e'$ generated by $f(x^d)$ is
\begin{align*}
    \begin{pmatrix}
    b_0&b_1&b_2&...&b_n&0&0&...&0\\
    0&b_0&b_1&b_2&...&b_n&0&...&0\\
    0&0&b_0&b_1&b_2&...&b_n&0&...\\
    \ddots&\ddots&\ddots&\ddots&\ddots&\ddots&\ddots&\ddots&\ddots
    \end{pmatrix}_{(e'-n)\times e'}.
\end{align*}
Now let $h'(x)=c_0+c_dx^d+c_{2d}x^{2d}+...+c_{sd}x^{sd}$, where $sd||n$. Then a parity check matrix of $\langle f(x^d) \rangle$ is given by
\begin{align*}
    \begin{pmatrix}
    c_{sd}&c_{(s-1)d}&c_{(s-2)d}&...&c_0&0&0&...&0\\
    0&c_{sd}&c_{(s-1)d}&c_{(s-2)d}&...&c_0&0&...&0\\
    0&0&c_{sd}&c_{(s-1)d}&c_{(s-2)d}&...&c_0&0&...\\
    \ddots&\ddots&\ddots&\ddots&\ddots&\ddots&\ddots&\ddots&\ddots
    \end{pmatrix}_{n\times e'},
\end{align*}
Note that a generator matrix $G$ of $C$ is given by
\begin{align*}
    \begin{pmatrix}
    b_0&0&...&b_1&0&...&b_n&0&0&...\\
    0&b_0&0&...&b_1&0&...&b_n&0&...\\
    0&0&b_0&0&...&b_1&0&...&b_n&...\\
    \ddots&\ddots&\ddots&\ddots&\ddots&\ddots&\ddots&\ddots&\ddots&\ddots
    \end{pmatrix}_{(e-nd)\times e}.
\end{align*}
It follows that a parity check matrix of $C$ is given by
\begin{align*}
    \begin{pmatrix}
    c_{sd}&0&...&c_{(s-1)d}&0&...&c_{0}&0&0&...\\
    0&c_{sd}&0&...&c_{(s-1)d}&0&...&c_{0}&0&...\\
    0&0&c_{sd}&0&...&c_{(s-1)d}&0&...&c_{0}&...\\
    \ddots&\ddots&\ddots&\ddots&\ddots&\ddots&\ddots&\ddots&\ddots&\ddots
    \end{pmatrix}_{(e-nd)\times e}.
\end{align*}
Note that the parity check matrix of $C$ can also be written as 
\begin{align*}
    \begin{pmatrix}
    c_{sd}&c_{sd-1}&...&c_{(s-1)d}&...&c_{0}&0&0&...\\
    0&c_{sd}&c_{sd-1}&...&c_{(s-1)d}&...&c_{0}&0&...\\
    0&0&c_{sd}&c_{sd-1}&...&c_{(s-1)d}&...&c_{0}&...\\
    \ddots&\ddots&\ddots&\ddots&\ddots&\ddots&\ddots&\ddots&\ddots
    \end{pmatrix}_{(e-nd)\times e}.
\end{align*}
Hence, for all $c_j$ where $j\notdivides d$, we have $c_j=0$. Therefore, $h(x)=c_0+c_1x+x_2x^2+...+c_mx^m=c_0+c_dx^d+c_{2d}x^{2d}+...+c_{sd}x^{sd}=h'(x)$ can be written as $h''(x)=c_0+c_dx+c_{2d}x^{2}+...+c_{sd}x^{s}$. Therefore, $\gcd(s_j)|e.$
\end{proof}

\begin{theorem}
For polynomial $f(x)=\sum_{a_j\not=0}a_jx^{s_j}\in \mathbb{F}_q[x]$, where at least one $s_j$  is a power of $q$, $\gcd(s_1,s_2,...)|\ord(f)$.
\end{theorem}
\begin{proof}
Let $s_j=q^{r_j}\cdot c_j$, and then $\gcd(s_1,s_2,...)=q^{\gcd(r_1,r_2,...)}$ since there exists at least one $s_j$ that is a power of $q$. Note that $f(x)$ can be written as
\begin{align*}
    f(x)&=\sum_{a_j\not=0}a_jx^{s_j}\\
    &=\left(\sum_{a_j\not=0}a_jx^{c_j\cdot \frac{r_j}{\gcd(r_1,r_2,...)}}\right)^{q^{\gcd(r_1,r_2,...)}} 
\end{align*}
Let $g(x)=\sum_{a_j\not=0}a_jx^{c_j\cdot \frac{r_j}{\gcd(r_1,r_2,...)}}=\prod p_i^{s_i}$, where $p_i$ is irreducible. Hence, $f(x)$ is given by
\begin{align*}
    f(x)&=\left(\sum_{a_j\not=0}a_jx^{c_j\cdot \frac{r_j}{\gcd(r_1,r_2,...)}}\right)^{q^{\gcd(r_1,r_2,...)}}\\
    &=g^{q^{\gcd(r_1,r_2,...)}}(x)\\
    &=\left(\prod p_i^{c_i}\right)^{ q^{\gcd(r_1,r_2,...)}}\\
    &=\prod p_i^{c_i\cdot q^{\gcd(r_1,r_2,...)}}.
\end{align*}
Let $q^t\geq \max(c_i\cdot q^{\gcd(r_1,r_2,...)}) \geq q^{\gcd(r_1,r_2,...)}$. Then $\gcd(s_1,s_2,...)=q^{\gcd(r_1,r_2,...)}|q^t$. Therefore, $\gcd(s_1,s_2,...)|\ord(f(x))=q^t\cdot \mbox{lcm}(\ord(p_i))$.
\end{proof}

\begin{theorem}
For any polynomial $f(x)\in\mathbb{F}_q[x]$, $\ord(f(x^i))=i\cdot\ord(f(x))$
\end{theorem}

\begin{proof}
We prove this result by proving $\Fq[x]/\langle x^e-1 \rangle \cong \Fq[x^i]/\langle x^{ei}-1 \rangle$.
Consider the map $\phi:\Fq[x] \xrightarrow[]{} \Fq[x^i]/\langle x^{ei}-1 \rangle$ defined by $\phi(f(x))=f(x^i)+\langle x^{ei}-1 \rangle$. It is a surjective ring homomorphism. Note that $\phi(x^e-1)=0$. Hence, $\langle x^e-1 \rangle \subseteq Ker(\phi)$. By the first isomorphism theorem,  $\displaystyle{ \Fq[x]/ Ker(\phi) \cong  \Fq[x^i]/\langle x^{ei}-1\rangle}$.  Since the sizes of the sets $\Fq[x]/\langle x^e-1 \rangle$ and $\Fq[x^i]/\langle x^{ei}-1 \rangle$ are the same and  $\langle x^e-1 \rangle \subseteq Ker(\phi)$, we must have  $\langle x^e-1\rangle = Ker(\phi)$. Hence the isomorphism of the quotient rings follows. Finally, this isomorphism implies the assertion on the order of the polynomials.
\end{proof}

We now introduce the notion of an expanded code and enlarged code that will be useful in our search method.
\begin{definition}[code expansion]
For a given cyclic code of length $n$ and dimension $k$ over $\Fq$, there exists a cyclic code of length $n\cdot i$ and dimension $k\cdot i$ over $\Fq$, for any positive integer $i$. In other words, for any cyclic code $C=\langle g(x) \rangle$ where $g(x)|x^n-1$,  with parameters $[n,k]$, we always have a corresponding cyclic code $C^{e}=\langle g(x^i)\rangle$ with parameters $[ni,ki]$ which is an ideal in $\Fq[x]/\langle x^{ni}-1 \rangle$.  We call $C^{e}$ an expanded code of $C$.
\end{definition}
%\ol{from my computational evidence, it is %correct} \hl{Clarify the quotient ring and  %prove it. If you can, then we have another %theorem.} \hlc[green]{cant resolve the min d %problem}

%\hlc[green]{what about the case in $\Fq[x^i]/\langle x^{ni}-1 \rangle$? I wrote a similar argument below. }

%\hl{Is this a conjecture or a corollary of Thm 3.23?}
%\ol{Yes. This is the original idea I talked about. However, I got rid of the min d condition and changed the dimension.}
\begin{definition}[code enlargement]
For a given cyclic code $C=\langle g(x)\rangle$ of length $n$ and dimension $k$ over $\Fq$, there exists a cyclic code  $C^{E}=\langle g(x^i)\rangle$ of length $n\cdot i$, and dimension $k$ over $\Fq$, for any positive integer $i$. Here, $C$ is an ideal in $\Fq[x]/\langle x^{n}-1 \rangle$ and  $C^{E}$ is an ideal in $\Fq[x^i]/\langle x^{ni}-1 \rangle$.  We call $C^{E}$ an enlarged code of $C$.

\end{definition}

\noindent Using  results from \cite{gf7}, we have 

\begin{lemma}
Let the order of $x^n-1\in \Fq[x]$  be $e$ and the multiplicative order of $a$ modulo $q$ be $z$, then $\ord(x^n-a)=nz$.
\end{lemma}

\begin{definition}
Let $f(x)\in \Fq[x]$ be a nonzero polynomial. If $f(x)\not=0$, then the least positive integer $e$ for which $f(x)|x^e-a$ for some $a\in \Fq^{*}$, is called the semi order of $f$ and denoted by $\sord(f)$. If $f(0)=0$, then we first write $f(x)=x^hg(x)$ with $g(0)\not=0$, and define $\sord(f)=\sord(g)$.
\end{definition}
\begin{remark}
Note that for a given $n$ and $a\not=b$, 
$\gcd(x^n-a,x^n-b)=1$. Hence, there exists a unique semi order for any polynomial $f\in \Fq[x]$.
\end{remark}

%\hl{I commented out the following two conjectures. I tried to prove them but did not find one. I am also surprised that the results did not involve the degree of $f$. So, unless you can find a proof, I recommend not including them.}
%\ol{Yea, I was surprised too. However, I have a potential explanation for it: the degree of f is related to the order of f, and $q^e-1$ is a really large number, so it does make sense.}

\begin{comment}
We have a couple of conjectures related to the semi order.
\begin{conjecture}
For any (irreducible) polynomial $f \in \Fq[x]$, let $f|x^{\sord(f)}-a$ and $z$ be the multiplicative order of $a$ in $\Fq^{*}$, then $z |  q^e-1$, where $e=\ord(f)$.
\end{conjecture}
\hlc[green]{I think this conjecture is related to the fact that $\ord(f)=nz$, where $z$ is the multiplicative order of $a$ in $\Fq^{*}$. I guess shouldn't be too hard}

\ol{I found a very similar statement for order. Could u look at page 75 THM 3.3 and COROLLARY3.4 of the book introduction to Finite Fields? I'm confused about the part of splitting fields}
\begin{conjecture}
For any (irreducible) polynomial $f \in \Fq[x]$ and $f|x^{\sord(f)}-a$, $\sord(f) | (q^e-1)/z$, where $e=\ord(f)$ and $z$ is the multiplicative order of $a$ in $\Fq^{*}$.
\end{conjecture}
\end{comment}

\begin{theorem}
For any polynomial $f\in \Fq[x]$, the semiorder of $f$ is equal to the semiorder of its reciprocal $f^*$, i.e., $\sord(f)=\sord(f^*)$.
\end{theorem}
\begin{proof}
Let $f(x)|x^e-a$ and let $x^e-a=f(x)g(x)$. Then we have the following
\begin{align*}
    (x^e-a)^*&=f^*(x)g^*(x)\\
    x^e\cdot ((\frac{1}{x})^e-a)&=f^*(x)g^*(x)\\
    1-ax^e&=f^*(x)g^*(x)\\
    x^e-a^{-1}&=-a^{-1}f^*(x)g^*(x)
\end{align*}
Hence, $f(x)|x^e-a$ if and only if $f^*(x)|x^e-a^{-1}$, which implies that $\sord(f)=\sord(f^*)$. If $f(x)=0$, then let $f(x)=x^hg(x)$, where $g(0)\not=0$. Then we also have $\sord(g)=\sord(f)=\sord(f^*)=\sord(g^*)$.
\end{proof}

\begin{comment}
\begin{theorem}
Let the semiorder of $f(x)$ be $e$. Then we have a ring isomorphism 
\begin{align*}
    \Fq[x]/\langle f(x)\rangle &\cong  \Fq[x^e]/\langle x^{e}-1\rangle
    %\phi(f(x))&=f(x^z)
\end{align*}
\end{theorem}
\begin{proof}
Define a map $\phi: \Fq[x] \xrightarrow[]{} \Fq[x^e]/\langle x^{e}-1 \rangle$ by $\phi(x)=x^e+\langle x^{e}-1 \rangle$.\hl{This was all  copy-and-paste and was not the right thing for this proof. Please revisit.} \ol{Do you think it's a isomorphism? I feel like it's not even an isomorphism}

Note that the kernel of $\phi$ is $\{f(x)\in F_q[x] : \phi(f(x))\in \langle x^{e}-1 \rangle \}$, so it follows that for all $f(x)\in Ker(\phi)$, $\phi(f(x))=f(x^z)+\langle x^{e}-1 \rangle \in \langle x^{e}-1 \rangle$, i.e. $x^{e}-1 | f(x^i)$. It's not hard to see that for all $Ker(\phi)=\langle x^n-a \rangle$. Now we claim that $\phi$ is surjective. For any $f(x^z)+\langle x^{e}-1 \rangle\in F_q[x^z]/\langle x^{e}-1 \rangle$, we have $\phi(f(x))=f(x^z)+\langle x^{e}-1 \rangle$, so $\phi$ is surjective.

Therefore, we have the following isomorphism.
\begin{align*}
    F_q[x]/\langle x^n-a\rangle &\cong  F_q[x^z]/\langle x^{e}-1\rangle\\
    \phi(f(x))&=f(x^z)
\end{align*}
\end{proof}
\end{comment}

\section{Quasi polycyclic Codes}
\begin{definition}
A linear code $C$ is said to be an $r$-generator quasi-polycyclic (QP) code of index $\ell$ if it has a generator matrix of the  form
\begin{align*}
    \begin{pmatrix}
    G_{11} & G_{12}&...&G_{1\ell}\\
    G_{21} & G_{22}&...&G_{2\ell}\\
    \vdots & \vdots & &\vdots\\
    G_{r1}&G_{r2}&...&G_{r\ell}
    \end{pmatrix}_{rm\times n}
\end{align*}
where each $G_{ij}$ is a generator matrix of a polycyclic code.  The special case of a 1-generator quasi-polycyclic code with $\ell$ blocks has a generator matrix of the form
\begin{align*}
    \begin{pmatrix}
    G_{11} & G_{12}&...&G_{1\ell}
    \end{pmatrix}_{m\times n}
\end{align*}
\end{definition}
Focusing on  1-generator quasi-polycyclic codes, we show that we can adopt and generalize the ASR search algorithm that has been used extensively for 1-generator QT codes (\cite{ASR}, \cite{GenASR}). First, we prove a lemma which is a generalization of the analogous result from cyclic codes.

\begin{lemma}
Let  $C=\langle g(x) \rangle$ be a polycyclic code of length $n$ associated with $x^n-a(x)$ so that $x^n-a(x)=g(x)h(x)$ (hence $g(x)$ is the standard generator polynomial and $h(x)$ is the check polynomial). If  $f(x)$ is relatively prime with $h(x)$,  then $C=\langle g(x)\rangle=\langle g(x)f(x)\rangle$.
\end{lemma}
\begin{proof}
 The inclusion $\langle g(x)f(x)\rangle \subseteq \langle g(x)\rangle$ is obvious. Now let $g(x)\cdot q(x)\in\langle g(x)\rangle$. Since $\gcd(h(x),f(x))=1$, there exist $A(x), B(x)$ such that  $A(x)h(x)+B(x)f(x)=1$. Then we have
\begin{align*}
    A(x)h(x)+B(x)f(x)&=1\\
    g(x)A(x)h(x)+g(x)B(x)f(x)&=g(x)\\
    A(x)(x^n-a(x))+g(x)B(x)f(x)&=g(x)
\end{align*}
Hence, $g(x)B(x)f(x)\equiv g(x) \pmod{x^n-a(x)}$, so $g(x)\cdot q(x)\equiv g(x)B(x)f(x)\cdot q(x)\equiv g(x)f(x)\cdot B(x)q(x) \pmod{x^n-a(x)}$. Therefore, $\langle g(x)\rangle\subseteq\langle g(x)f(x)\rangle$, and $C=\langle g(x)\rangle=\langle g(x)f(x)\rangle$.
\end{proof}

\begin{theorem}[lower bound on minimum distance]
Let   $C=\langle g(x)\rangle$  be a polycyclic code of length $n$ associated with $a(x)$ and the standard generator polynomial $g(x)$ so that $x^n-a(x)=g(x)h(x)$.
Let $P=\langle g(x)f_1(x),g(x)f_2(x),g(x)f_3(x),...,g(x)f_{\ell}(x) \rangle$ be a QP code where 
$\gcd(f_i(x),h(x))=1$ for $i=1,2,..,\ell$. Then $P$ has minimum distance $D\geq l\cdot d$, where $d$ is the minimum distance of $C$.
\end{theorem}
\begin{proof}
Note that if $q(x)\cdot g(x)f_i(x)=0$, i.e., the $i^{th}$ block is zero, then $h(x)\mid q(x)\cdot f_i(x)$. Since $\gcd(f_i(x),h(x))=1$, it follows that $h(x)\mid q(x)$. In other words, for $i=1,2,..,\ell$,  $q(x)\cdot g(x)f_i(x)=0$. Hence, one block ($\langle g(x)f_i(x)\rangle$) of $P$ is zero if and only if all blocks are. Also note that $\langle g(x)f_i(x)\rangle=\langle g(x)\rangle$ and every nonzero codeword in each block has weight greater than or equal to $d$, so any nonzero codeword in $D$ has weight greater than or equal to $d\cdot \ell$.
\end{proof}

%\hlc[green]{(The comment below) I looked at the MT paper again, and I believed that it's slightly different than what we stated in the MT paper. Shall we include it still?}

\begin{comment}
The following theorem is easy to prove.
\hlc[green]{Do we need to repeat this thm again? Cuz we have smth similar in the MT paper}
\begin{theorem}
Let $P=\langle g_1(x),g_2(x),g_3(x),...,g_{\ell}(x) \rangle$ be a QP code and let $H_i$ be a generator matrix of the dual of $\langle g_i(x) \rangle$. Let $PD'$ be the code with generator matrix $(H_1,H_2,H_3,...,H_{\ell})$. Then $PD'$ is contained in the dual of $P$. In particular, if the dimension of $PD'$ is equal to the dimension of the dual of $P$, then $PD'$ is the dual of $P$. If $g_j(x)$ is of degree 0, then $H_1\bigoplus H_2\bigoplus H_3\bigoplus ... \bigoplus H_{\ell}$ is the dual of $P$.
\end{theorem}
%\hlc[orange]{the reversibility theorem also works here}
\end{comment}

\section{Quantum Codes}
\subsection{Quantum polycyclic Codes}
  The CSS construction requires two linear codes $C_1$ and $C_2$ such that $C_2^{\perp}\subseteq C_1$. When we work with cyclic, constacylic, or polycyclic codes that are generated by a single polynomial, it is useful to recall that  $\langle g_2\rangle \subseteq \langle g_1\rangle $ if and only if $g_1|g_2$. Hence, a polycyclic code $C_2$ generated by $g_2$ is a subcode of polycyclic code $C_1$ generated by $g_1$. Moreover, we can also think of $C_2$ as $C'^{\perp}_2$, so it follows that $C'^{\perp}_2\subseteq C_1$. Hence, by CSS construction, we can generate quantum CSS codes over GF($q^2$).

\begin{lemma}
Fix $n,i\in \mathbb{Z}$, and let $a_1, a_2, b \in \Fq^{*}$ be nonzero with $a_1\not=a_2$. Then $\gcd(x^n-a_1x^i-b,x^n-a_2x^i-b)=1$
\end{lemma}
\begin{proof}
Note that $\gcd(x^n-a_1x^i-b,x^n-a_2x^i-b)=\gcd(x^n-a_1x^i-b,x^n-a_1x^i-b-(x^n-a_2x^i-b))=\gcd(x^n-a_1x^i-b,(a_2-a_1)x^i)$. Also, $x^n-a_1x^i-b$ is  not a multiple of $x$, so $\gcd(x^n-a_1x^i-b,(a_2-a_1)x^i)=1$.
\end{proof}

\begin{lemma}
Fix $n,i\in \mathbb{Z}$, and let $a,b_1, b_2\in \Fq$, with $b_1\not=b_2$. Then $\gcd(x^n-ax^i-b_1,x^n-ax^i-b_2)=1$.
\end{lemma}
\begin{proof}
Let $\delta$ be a root of $x^n-ax^i-b_1=0$. Then it follows that 
\begin{align*}
    \delta^n-a\delta^i-b_2=(\delta^n-a\delta^i-b_1)+b_1-b_2=b_1-b_2\not=0.
\end{align*}
Therefore, $x^n-ax^i-b_1$ and $x^n-ax^i-b_2$ share no common roots. Therefore, $\gcd(x^n-ax^i-b_1,x^n-ax^i-b_2)=1$.
\end{proof}

\begin{theorem}
Let $n$ and $i$ be positive integers with $i<n$. For two distinct trinomials $x^n-ax^i-b$ and $x^n-cx^i-d$, we have $\gcd(x^n-ax^i-b,x^n-cx^i-d)$ is either 1 or a binomial of degree $\gcd(n,i)$.
\end{theorem}
\begin{proof}
Note that $\gcd(x^n-ax^i-b,x^n-cx^i-d)=\gcd(x^n-ax^i-b,x^n-ax^i-b-(x^n-cx^i-d))=\gcd(x^n-ax^i-b,(c-a)x^i+(d-b))$. Using a similar argument, we have the following,
\begin{align*}
    \gcd(x^n-ax^i-b,x^n-cx^i-d)&=\gcd(x^n-ax^i-b,(c-a)x^i+(d-b))\\
    &=\gcd(x^n-ax^i-b,x^i+(d-b)(c-a)^{-1})\\
    &=\gcd(x^n-ax^i-b+ a\cdot(x^i+(d-b)(c-a)^{-1}),x^i+(d-b)(c-a)^{-1})\\
    &=\gcd(x^n+a(d-b)(c-a)^{-1}-b,x^i+(d-b)(c-a)^{-1})
\end{align*}
Therefore, by Theorem 4.1 in \cite{mt} $\gcd(x^n-ax^i-b,x^n-cx^i-d)$ is either 1 or a binomial of degree $\gcd(n,i)$.
\end{proof}

\begin{lemma}
For any $x^n-ax^i-b\not=x^n-cx^i-d\in \mathbb{F}_3[x]$, we have  $\gcd(x^n-ax^i-b,x^n-cx^i-d)=1$.
\end{lemma}
\begin{proof}
Given Lemma 5.1 and Lemma 5.2, we only need to prove $\gcd(x^n-x^i-2,x^n-2x^i-1)=1$. Note that $\gcd(x^n-x^i-2,x^n-2x^i-1)=\gcd(x^n-x^i-2,x^i-1)=\gcd(x^n,x^i-1)$. Also, $x^i-1$ is obviously not a multiple of $x$, so $\gcd(x^n-x^i-2,x^n-2x^i-1)=\gcd(x^n,x^i-1)=1$.
\end{proof}

\begin{theorem} 
If $\gcd(n,q)=\gcd(m,q)=1$, then  $x^n-a|x^m-b$ over $\Fq$ if and only if $n|m$ and $b=a^{m/n}$.
\end{theorem}
\begin{proof}
Suppose $x^n-a|x^m-b$. Let $m=kn+r$, where $r<n$.  Let $\delta$ be a root of $x^n-a$ and let $\alpha$ be a primitive $n^{th}$ root of unity. Then  for any $i\in \mathbb{Z}^{+}$, we have 
\begin{align*}
    (\delta\cdot \alpha^i)^m-b&=0\\
    (\delta\cdot \alpha^i)^{kn+r}-b&=0\\
    (\delta^{kn}\cdot \alpha^{ikn})(\delta^r\cdot \alpha^{ir})-b&=0\\
    a^k(\delta^r\cdot \alpha^{ir})-b&=0\\
    \delta^r\cdot \alpha^{ir}&=ba^{-k}
\end{align*}
Since $i$ could be any positive integer, we have the following:
\begin{align*}
    \delta^r\cdot \alpha^{r}&=ba^{-k}\\
    \delta^r\cdot \alpha^{2r}&=ba^{-k}
\end{align*}
Hence, $\alpha^{r}=\alpha^{2r}$ implies $r=0$, so $m=nk$ and $b=a^k$.

Now suppose $n|m$ and $b=a^{m/n}$, and let $m=nk$. Let $\delta$ be a root of $x^n-a=0$ and $\alpha$ be a primitive $n^{th}$ root of unity. Then for any $i\in \mathbb{Z}^{+}$, we have 
\begin{align*}
    (\delta\cdot \alpha^i)^m-b&=(\delta\cdot \alpha^i)^{nk}-b\\
    &=\delta^{nk}\cdot \alpha^{ink}-b\\
    &=a^k-b=0.
\end{align*}
Therefore, every root of $x^n-a$ is also a root of $x^m-b$, so $x^n-a|x^m-b$.
\end{proof}

%\hl{Binary example is not the best. Better to use another field and take and example where not both $a$ and $b$ are 1.}
\begin{ex}
Over $\mathbb{F}_5$, let $n=7,a=2,m=21,b=3$, then we have $\gcd(7,5)=\gcd(21,5)=1$, $7|21$ and $2=3^{21/7}$. Therefore, $x^7-2|x^{21}-3.$ Also, we know $x^2-2|x^4-1$ over GF(3) since $x^4-1=(x^2-2)(x^2+2)$. It is easy to verify that $\gcd(2,3)=\gcd(4,3)=1$, $2|4$ and $2^{4/2}=1$.
\end{ex}

\begin{theorem}[generalized]
Over $GF(q)$, where $q$ is a power of a prime $p$, let $n=p^{n_1}\cdot n_2$, $m=p^{m_1}\cdot m_2$, $a'=a^{p^{-n_1}}$ and $b'=b^{p^{-m_1}}$. Then $x^n-a|x^m-b$ if and only if $n|m$ and $b'=(a')^{m_2/n_2}$.
\end{theorem}
\begin{proof}
Suppose $x^n-a|x^m-b$. Let $n=p^{n_1}\cdot n_2$ and $m=p^{m_1}\cdot m_2 $. Since $x^n-a|x^m-b$, it follows that $n_1\leq m_1$. We have 
\begin{align*}
    x^n-a&=(x^{n_2}-a')^{p^{n_1}}\\
    x^m-b&=(x^{m_2}-b')^{p^{m_1}},
\end{align*}
where $(a')^{p^{n_1}}=a$ and $(b')^{p^{m_1}}=b$. We also have $x^{n_2}-a'|x^{m_2}-b'$ since all irreducible factors of $x^{n_2}-a'$ are of multiplicity 1 and have to be contained among the irreducible factors of $x^{m_2}-b'$. Let $m_2=kn_2+r$, where $r<n_2$.  Let $\delta$ be a root of $x^{n_2}-a'$ and let $\alpha$ be a primitive $n_{2}^{th}$ root of unity. Then for any $i\in \mathbb{Z}^{+}$, we have 
\begin{align*}
    (\delta\cdot \alpha^i)^{m_2}-b'&=0\\
    (\delta\cdot \alpha^i)^{kn_2+r}-b'&=0\\
    (\delta^{kn_2}\cdot \alpha^{ikn_2})(\delta^r\cdot \alpha^{ir})-b'&=0\\
    a'^k(\delta^r\cdot \alpha^{ir})-b'&=0\\
    \delta^r\cdot \alpha^{ir}&=b'(a')^{-k}
\end{align*}
Since $i$ could be any positive integer, we have 
\begin{align*}
    \delta^r\cdot \alpha^{r}&=b'(a')^{-k}\\
    \delta^r\cdot \alpha^{2r}&=b'(a')^{-k}
\end{align*}
Hence, $\alpha^{r}=\alpha^{2r}$ implies $r=0$, so $m_2=n_2k$, $n|m$ and $b'=(a')^k$.

Now suppose $n_2|m_2$ and $b'=a'^{m_2/n_2}$. Note that $n=p^{n_1}\cdot n_2|m=p^{m_1}\cdot m_2$, so it follows that $n_1\leq m_1$ and $n_2|m_2$.
Let $m_2=n_2k$ and let $\delta$ be a root of $x^n_2-a'=0$ and $\alpha$ be a primitive $n_2^{th}$ root of unity. Then for any $i\in \mathbb{Z}^{+}$, we have 
\begin{align*}
    (\delta\cdot \alpha^i)^{m_2}-b'&=(\delta\cdot \alpha^i)^{n_2k}-b'\\
    &=\delta^{n_2k}\cdot \alpha^{in_2k}-b'\\
    &=(a')^k-b'=0.
\end{align*}
Therefore, every root of $x^{n_2}-a'$ is also a root of $x^{m_2}-b'$, so $x^{n_2}-a'|x^{m_2}-b'$. Hence, $(x^{n_2}-a')^{p^{n_1}}=x^n-a|x^m-b=(x^{m_2}-b')^{p^{m_1}}$.
\end{proof}

\begin{ex}
Over $GF(7)$, let $n=7^1\cdot2=14$, $m=7^2\cdot 8=392$, $a=2$, $b=2$ and consider $x^n-a=x^{14}-2$ and $x^m-b=x^{392}-2$. It's not hard to see that $a'=a^{7^{-1}}=2,b'=b^{7^{-2}}=2$.  Given that $n=1|2=m$ and $4=b'=(a')^{8/2}=4,$ it follows that $x^{14}-2|x^{392}-2$. 

Also over GF(7), since $x^{15} - 4=(x^{12} + 2x^9 + 4x^6 + x^3 + 2)(x^3-2)$, we have $x^3-2 |x^{15}-4$. Letting $a=2, b=4$, we can verify that $15=7^0\cdot 15$,$3=7^0\cdot 3$ and $a'=2^{-1}=4,b'=4^{-1}=2$. Thus, we have $3|15$ and $b'=2=(a')^{15/3}=4^5=2.$ 
\end{ex}

 This theorem helps us simplify the search process for quantum codes. For example, if we search for quantum codes of length 100, we normally find all possible $g(x)|x^{100}-a$ and $f(x)g(x)|x^{100}-a$ and then construct CSS codes using $C_1=\langle g(x)\rangle$ and $C_2=\langle g(x)f(x)\rangle^{\perp}$, which would take a long time. However, with this theorem, we can find some $x^n-b|x^{100}-a$, and then find all possible $g(x)|x^n-b$ and $f(x)g(x)|x^n-b$.

Now let $C_1=\langle g(x)\rangle$ and $C_2=\langle g(x)f(x)\rangle^{\perp}$ both be of length $n$. Then using the idea of code expansion, we can get $C_1'=\langle g(x^{100/n})\rangle$ and $C_2'=\langle g(x^{100/n})f(x^{100/n})\rangle^{\perp}$ both of length $100$. Finally, we construct a CSS code using $C_1'$ and $C_2'$, which is a quantum code of length 100.

\section{New Codes}
We obtained a record-breaking polycyclic code, $[92, 4, 82]_{17}$, from the trinomial $x^{92}-x^{14}-12$ with check polynomial $h = x^4 + 2x^3 + 16x^2 + 12x + 12$ which has a higher minimum distance than the $[92,4,81]$-code given in Table 9 of  \cite{GF17&19}.

\subsection{Polycyclic Codes with the Same Parameters as BKLCs}
There are many polycyclic codes with the same parameters are BKLCs. They generally occur when $n$ is small. A lot of them have additional desirable properties such as reversibility and self-duality. Also, some of the codes in the table below have simpler constructions than the ones given in database \cite{database}. The polynomials in the tables below are represented by coefficients in ascending order of exponent. For example, $[1011]$ over $\mathbb{F}_{2}$ denotes $1 + x^2 +x^3$. In addition, for codes over $\mathbb{F}_{q}$ with $q > 10$, $A$ denotes $10$, $B$ denotes $11, ... ,$ and $G$ denotes 16.

\label{tab:1}       % Give a unique label
\newpage
\begin{longtable}{p{2.5cm}p{1cm}p{1cm}p{1cm}p{1cm}p{4.5cm} }

\caption{\small{Polycyclic codes with the same parameters as BKLCs}}
\\\hline\noalign{\smallskip}
$[n,k,d]_{q}$  &$n$& $i$ & $a$ & $b$ & Polynomial  \\
\noalign{\smallskip}\hline\noalign{\smallskip}
     $[19, 12, 4]_{2}$$^{\star}$$^{\circ}$ &19 & 18 & 1 &1 & [101010011] \\
  [.5ex] 
  $[21, 14, 4]_{2}$$^{\star}$$^{\circ}$ &21 & 3 & 1 &1 & [11001011] \\
  [.5ex] 
  $[23, 18, 3]_{2}$$^{\star}$$^{\circ}$ &23 & 12 & 1 &1 & [101001] \\
  [.5ex] 
  $[24, 19, 3]_{2}$$^{\star}$$^{\circ}$ &24 & 22 & 1 &1 & [101111] \\
  [.5ex] 
  $[14, 6, 6]_{3}$$^{\star}$$^{\circ}$ &14 & 7 & 1 &1 & [101201101] \\
  [.5ex]
  $[17, 8, 6]_{3}$$^{\star}$$^{\circ}$ &17 & 10 & 1 &1 & [2010121201] \\
  [.5ex]
  $[27, 15, 7]_{3}$$^{\star}$$^{\circ}$ &27 & 25 & 1 &1 & [1002012222121] \\
  [.5ex]
  $[16, 6, 7]_{3}$$^{\star}$$^{\circ}$ &16 & 13 & 1 &2 & [20022011201] \\
  [.5ex]
  $[18, 9, 6]_{3}$$^{\star}$$^{\circ}$ &18 & 15 & 1 &2 & [2101011121] \\
  [.5ex]
  $[14, 6, 6]_{3}$$^{\star}$$^{\circ}$ &14 & 7 & 2 &1 & [2101011121] \\
  [.5ex]
   $[17, 8, 6]_{3}$$^{\star}$$^{\circ}$ &17 & 1 & 2 &1 & [2101011121] \\
  [.5ex]
  $[15, 8, 5]_{3}$$^{\star}$$^{\circ}$ &15 & 10 & 2 &2 & [2101011121] \\
  [.5ex]
   $[22, 7, 10]_{4}$$^{\star}$$^{\circ}$ &22 & 11 & 1 &1 & $[\alpha^{2}0\alpha^{2}110\alpha^{2}1\alpha\alpha^{2}10\alpha^{2}\alpha^{2}101]$ \\
  [.5ex]
  $[20, 9, 8]_{4}$$^{\star}$$^{\circ}$ &20 & 15 & 1 &$\alpha$ & $[\alpha^{2}\alpha^{2}\alpha0\alpha\alpha^{2}\alpha^{2}1\alpha\alpha^{2}\alpha^{2}1]$ \\
  [.5ex]
    $[29, 19, 6]_{4}$$^{\star}$$^{\circ}$ &20 & 15 & 1 &$\alpha^{2}$ &$[\alpha^{2}011\alpha^{2}\alpha^{2}101\alpha^{2}1]$ \\
  [.5ex]
  $[20, 9, 8]_{4}$$^{\star}$$^{\circ}$ &20 & 15 & $\alpha$&1  &$[\alpha^{2}1\alpha^{2}0\alpha\alpha\alpha^{2}\alpha^{2}\alpha^{2}\alpha^{2}11]$ \\
  [.5ex]
  $[22, 13, 6]_{4}$$^{\star}$$^{\circ}$ &22 & 19 & $\alpha$ & $\alpha$ &$[1\alpha\alpha^{2}01\alpha^{2}1101]$ \\
  [.5ex]
  $[28, 13, 6]_{4}$$^{\star}$$^{\circ}$ &22 & 19 & $\alpha$ & $\alpha^{2}$ &$[1\alpha^{2}1\alpha^{2}\alpha^{2}\alpha^{2}11011]$ \\
  [.5ex]
    $[20, 15, 4]_{5}$$^{\star}$$^{\circ}$ &20 & 5 & $2$ & $2$ &$[130324]$ \\
  [.5ex]
  $[30, 16, 9]_{5}^{\circ}$ &30 & 1 & $4$ & $2$ &$[142132431343042]$ \\
  [.5ex]
\noalign{\smallskip}\hline
\end{longtable}
\hspace{-.4cm}$\star$ delineates optimal codes;\\ 
\hspace{-3.4cm}$^\circ$ delineates simpler construction codes.\\

\begin{longtable}{p{2.5cm}p{1cm}p{1cm}p{1cm}p{1cm}p{2cm} }

\caption{\small{Reversible polycyclic codes with same parameters as BKLCs}}
\\\hline\noalign{\smallskip}
$[n,k,d]_{q}$  &$n$& $i$ & $a$ & $b$ & Polynomial 
\\\noalign{\smallskip}\hline\noalign{\smallskip}
     $[10, 8, 2]_{2}^{\star}$ &10 & 2 & 1 &1 & [111] \\
  [.5ex] 
  $[14, 12, 2]_{2}^{\star}$ &14 & 13 & 1 &1 & [111] \\
  [.5ex] 
  $[20, 18, 2]_{2}^{\star}$ &20 & 1 & 1 &1 & [111] \\
  [.5ex] 
  $[32, 30, 2]_{2}^{\star}$ &14 & 13 & 1 &1 & [111] \\
  [.5ex] 
  $[10, 9, 2]_{3}^{\star}$ &10 & 7 & 1 &1 & [11] \\
  [.5ex] 
  $[10, 9, 2]_{3}^{\star}$ &10 & 7 & 1 &1 & [11] \\
  [.5ex] 
  $[42, 39, 2]_{3}^{\star}$ &42 & 9 & 1 &2 & [101] \\
  [.5ex] 
\noalign{\smallskip}\hline
\end{longtable}
\hspace{-.4cm}$\star$ delineates optimal codes;\\ 
\hspace{-3.4cm}$^\circ$ delineates simpler construction codes.\\

\subsection{Two-block quasi-polycyclic codes}

When searching QP codes, we are interested in six characteristics: reversibility, self-orthogonality, self-duality, record breaking minimum distance, and best known minimum distance. Below we list tables of codes with either record breaking minimum distances or two or more other characteristics. These are all 1-generator QP codes with generators in the form given in Theorem 7.2. When the degree of $g(x)$ is large, we use the check polynomial $h(x)$ instead.

\begin{table}[h]
\resizebox{1\textwidth}{!}{\begin{minipage}{\textwidth}
\caption{Self-Orthogonal QP Codes with Best Known Parameters}

\label{tab-2}
\begin{tabular}{p{2cm}p{.3cm}p{.3cm}p{.3cm}p{2.9cm}p{2.9cm}p{2.9cm}}
\hline\noalign{\smallskip}
$[m,k,d]_{q}$  & $i$ & $a$ & $b$ & $h$ & $f1$ & $f2$\\
\noalign{\smallskip}\hline\noalign{\smallskip}
     $[124, 6, 62]_{2}$ &7 & $1$ & $1$ & $[1011011]$  & $[000001]$ & $[000001]$\\
  [.5ex] 
  $[128, 6, 64]_{2}$ &6 & $1$ & $1$ & $[1100001]$  & $[000001]$ & $[000001]$\\
  [.5ex] 
   $[28, 6, 15]_{3}$ &7& $1$ & $1$ & $[1201]$ & $[20122]$ &$[210121]$  \\
  [.5ex]
   $[40, 8, 21]_{3}$ &10& $1$ & $1$ &$[121202011]$ & $[22011221]$ &$[2002222]$ \\
  [.5ex]
\noalign{\smallskip}\hline
\end{tabular}
\end{minipage}}
\end{table}

\begin{table}[h]
\resizebox{1\textwidth}{!}{\begin{minipage}{\textwidth}
\caption{QP Codes with Best Known Parameters}

\label{tab-3}
\begin{tabular}{p{2cm}p{.3cm}p{.3cm}p{.3cm}p{2.9cm}p{2.9cm}p{2.9cm}}
\hline\noalign{\smallskip}
$[n,k,d]_{q}$  & $i$ & $a$ & $b$ & $h$ & $f1$ & $f2$\\
\noalign{\smallskip}\hline\noalign{\smallskip}
  $[124, 6, 62]_{2}$ &7 & $1$ & $1$ & $[1011011]$  & $[000001]$ & $[000001]$\\
  [.5ex] 
  $[128, 6, 64]_{2}$ &6 & $1$ & $1$ & $[1100001]$  & $[000001]$ & $[000001]$\\
  [.5ex] 
   $[28, 6, 15]_{3}$ &7& $1$ & $1$ & $[1201]$ & $[20122]$ &$[210121]$ \\
  [.5ex]
   $[40, 8, 21]_{3}$ &10& $1$ & $1$ &$[121202011]$ & $[22011221]$ &$[2002222]$ \\
  [.5ex]
\noalign{\smallskip}\hline
\end{tabular}
\end{minipage}}
\end{table}

\begin{table}[h]
\resizebox{1\textwidth}{!}{\begin{minipage}{\textwidth}
\caption{QP Codes that are Self-Orthogonal and Reversible}

\label{tab-4}
\begin{tabular}{p{2cm}p{.3cm}p{.3cm}p{.3cm}p{2.9cm}p{2.9cm}p{2.9cm}}
\hline\noalign{\smallskip}
$[m,k,d]_{q}$  &$i$&$a$&$b$& $h$ &$f1$ & $f2$  \\
\noalign{\smallskip}\hline\noalign{\smallskip}
     $[40, 8, 8]_{2}$ &10 & $1$ & $1$ &$[110111011]$  & $[1001001]$ & $[1001001]$\\
  [.5ex] 
  $[42, 15, 4]_{2}$ &6 &$1$ & $1$ & $[1001001000001001]$  & $[100111101010101]$ & $[100111101010101]$\\
  [.5ex]
  $[44, 12, 10]_{2}$ &11 & $1$ & $1$ &$[1001011101001]$  & $[0000110101]$ & $[0000110101]$\\
  [.5ex]
   $[44, 12, 10]_{2}$ &11 &$1$ & $1$ & $[1001011101001]$  & $[0000110101]$ & $[0000110101]$\\
  [.5ex]
  
\noalign{\smallskip}\hline
\end{tabular}
\end{minipage}}
\end{table}

\subsection{Quantum Codes Constructed from polycyclic Codes}
Compared to classical coding theory, the field of quantum coding theory is still relatively young. The idea of quantum error correction codes was first introduced in \cite{Quantumoriginal1} and \cite{Quantumoriginal2}. A construction method (Quantum Error Correcting Code (QECC) is proposed in \cite{Quantumoriginal3}. However, amounts of methods to construct new QECCs are relatively complicated. A very direct way of using classical error correction codes to construct new QECCs is called the CSS construction\cite{Quantumoriginal3}. It requires two 
linear codes $C_1$ and $C_2$ such that $C_2^{\perp}\subseteq C_1$. Therefore, self-dual, self-orthogonal, and dual-containing codes are used frequently. For example, if $C_1$ is a self-dual, self-orthogonal, or dual-containing codes code, then we can construct a CSS quantum code using $C_1$ alone since either $C_1^{\perp}\subseteq C_1$ or $C_1^{\perp}\supseteq C_1$. Using the CSS construction, we obtained a good number of record breakers and other codes whose parameters do not appear in the literature. In addition, the record-breaking codes we discovered have  simpler constructions than those within the literature. A lot of codes in  the literature are constructed by indirect ways by first considering an extension ring then mapping back to the ground field. It is more desirable to obtain codes with the same even better parameters by directly working on the ground field.  

\begin{table}[h]
\resizebox{1\textwidth}{!}{\begin{minipage}{\textwidth}
\caption{\small{Quantum codes with record-breaking parameters constructed from polycyclic codes}}
\label{tab-5}
\begin{tabular}{p{2.3cm}p{2.3cm}p{3.6cm}p{3.6cm}p{.7cm}}
\hline\noalign{\smallskip}
$[[n,k,d]]_{q^2}$& $t$ &$g_1$& $h_2$ & Ref.\\
\noalign{\smallskip}\hline\noalign{\smallskip}
$[[36,4,8]]_{3^2}$ & $x^{36} - 2x^{2} - 1$ & $ [ 1 0 0 2 2 2 1 2 0 2 1 2 1 0 0 2 1 ]$ & $ [ 1 0 0 1 2 1 1 1 0 1 1 1 1 0 0 1 1 ]$ & \cite{src30}\\
$[[36,6,7]]_{3^2}$ & $x^{36}-x^{8}-1$ & $[ 2 1 2 0 0 1 2 0 1 0 2 1 1 2 1 1 ]$ & $[ 1 1 1 0 0 1 1 0 2 0 1 1 2 2 2 1 ]$ & \cite{src30}\\
 $[[60,32,5]]_{3^2}$ &$x^{60} - 2x^6 - 2$ &$[ 1 1 2 1 0 1 1 0 1 1 0 0 1 ]$ & $[ 2 0 2 2 0 0 0 2 0 2 2 1 0 2 0 2 1 ]$ & \cite{src30}\\
  $[[72,48,5]]_{3^2}$ & $x^{72}-x^{2} - 1$ & $[ 2 1 0 0 1 1 2 1 2 1 1 0 1 ]$ & $[ 2 2 0 0 1 2 2 2 2 2 1 0 1 ]$ & \cite{src29}\\
    $[[78,60,4]]_{3^2}$ & $x^{78} - 2x^{2} - 2$ & $[ 1 0 1 1 0 2 1 0 2 1 ]$ & $[ 2 0 2 1 0 2 2 0 1 1 ]$ & \cite{src5}\\
    $[[40,24,4]]_{5^2}$ & $x^{40} - 4x^4 - 2$  & $[2 2 2 4 3 1 1 2 1 ]$ & $[ 4 1 4 4 0 1 4 3 1 ]$ & \cite{src29}\\
    $[[40, 32, 3]]_{5^2}$ & $x^{40} - 2x^{2} - 3$ & $[ 2 0 2 2 1] $ & $[ 2 0 2 3 1 ]$ & \cite{src29}\\
    $[[45,33,4]]_{5^2}$ & $x^{45} - x^{18} - 1$  & $[ 4 3 1 0 0 0 1 ]$ & $[ 4 1 2 3 0 1 1 ]$ & \cite{src5}\\
    $[[60,48,4]]_{5^2}$ & $x^{60} - x^{10} - 2$ & $[ 1 4 4 3 1 3 1 ]$ & $[ 1 1 4 2 1 2 1 ]$ & \cite{src19}\\
    $[[72,60,4]]_{5^2}$ & $x^{72} - 2x^{12} - 4$ & $[ 4 0 3 0 1 3 1 ]$ & $[ 4 0 3 0 1 2 1 ]$ & \cite{src5}\\
    $[[80,56,4]]_{5^2}$ & $x^{80}-x-1$ & $[ 4 2 3 1 3 4 2 3 3 1 ]$ & $[ 2 3 0 1 3 3 0 0 0 2 3 1 4 4 2 1 ]$ & \cite{src14}\\
     $[[27,17,4]]_{7^2}$ & $x^{27} - 2x^3 - 3$ & $[ 1 1 3 1 3 1 ]$ & $[ 4 2 3 4 6 1 ]$ & \cite{src19}\\
    $[[16,10,3]]_{13^2}$ & $x^{16} - 2x - 4$ & $[ 1 4 9 1 ]$ & $[ 6 B B 1 ]$ &  \cite{src6}\\
    $[[18,10,4]]_{13^2}$ & $x^{18}-x-1$ & $[ 8 6 8 C 1 ]$ & $[ A 6 1 5 1 ]$ & \cite{src19}\\
    $[[24,16,4]]_{13^2}$ & $x^{24} - x^2 - 4$ & $[ C  C  2  6  1 ]$ & $[ C  1  2  7  1 ]$ & \cite{src29}\\
     $[[36,20,4]]_{13^2}$ & $x^{36} - x - 5$ & $[ 2  0  6  6  4  1 ]$ & $[ 3 A 1 1 5 5 3 C 5 0 5 1]$ & \cite{src26}\\
    $[[24,18,3]]_{17^2}$ & $x^{24} - x^{2} - 3$ & $[ C G 6 1 ]$ & $[5 G B 1]$ & \cite{src18}\\
    $[[36,30,3]]_{17^2}$ & $x^{36} - x - 1$ & $[ 3 7 9 1 ]$ & $[ 1 G F 1 ]$ & \cite{src18}\\
  [.5ex] 
\noalign{\smallskip}\hline
\end{tabular}
\end{minipage}}
\end{table}
%A-10, B-11, C-12, D-13, E-14, F-15, G-16

\newpage

\begin{longtable}{p{2.2cm}p{2.2cm}p{4.5cm}p{3.5cm}p{.7cm}}

\caption{\small{Quantum codes  with the best known parameters constructed from polycyclic codes}}
\\\hline\noalign{\smallskip}
$[[n,k,d]]_{q^2}$  &$t$& $g_1$ & $h_2$ & Ref.\\ 
\noalign{\smallskip}\hline\noalign{\smallskip}
  $[[36, 32, 2]]_{3^2}^{\circ}$ &  $x^{36} - x - 2$ & $[ 1 2 1 1 ]$ & $[ 1 1 ]$ & \cite{src14}\\
  [.5ex]
  $[[45, 21, 5]]_{3^2}^{\circ}$ &  $x^{45} - x^5 - 1$ & $[ 1 0 2 2 2 0 2 0 2 0 2 0 1 ]$ & $[ 1 2 0 2 1 1 2 1 2 0 2 1 1 ]$ & \cite{src5}\\
  [.5ex]
  $[[60, 36, 4]]_{3^2}^{\circ}$ &  $x^{60} - x^6 - 1$ & $[ 1 2 2 2 0 1 2 1 1 0 1 ]$ & $[ 1 2 2 0 0 2 2 2 1 2 2 1 2 0 1 ]$ & \cite{src5}\\
  [.5ex]
   $[[72, 66, 2]]_{3^2}^{\circ}$ & $x^{72} - 2x^{2} - 2$ & $[ 2 0 1 1 ]$ & $[ 1 0 2 1 ]$ & \cite{src18}\\
  [.5ex]
   $[[78, 60, 3]]_{3^2}^{\circ}$ & $x^{78} - 2x^2 - 1$ & $[ 2 2 1 2 2 1 2 1 1 ]$ & $[ 2 1 0 2 0 0 1 1 0 2 1 ]$ & \cite{src5}\\
   $[[25, 18, 3]]_{5^2}^{\circ}$ & $x^{25}-x^4-2$ & $[ 2 1 3 1 ]$ & $[ 4 0 0 2 1 ]$ & \cite{src32}\\
  [.5ex]
  $[[25, 22, 2]]_{5^2}^{\circ}$ & $x^{25}-x^3-1$ & $[ 3 3 1 ]$ & $[ 2 1 ]$ & \cite{src32}\\
  [.5ex]
  $[[40, 34, 2]]_{5^2}^{\circ}$ & $x^{40}-x^{2}-4$ & $[ 3 0 1 ]$ & $[ 3 0 2 0 1 ]$ & \cite{src6}\\
  [.5ex]
  $[[45, 33, 3]]_{5^2}^{\circ}$ & $x^{45}-3x-1$ & $[ 4 2 3 3 1 ]$ & $[ 1 0 0 0 0 1 0 3 1 ]$ & \cite{src5}\\
  [.5ex]
   $[[54, 46, 2]]_{5^2}^{\circ}$ & $x^{54}-x^2-1$ & $[ 3 0 1 ]$ & $[ 1 0 4 0 4 4 1 ]$ & \cite{src5}\\
  [.5ex]
   $[[49, 40, 3]]_{7^2}^{\circ}$ & $x^{49}-2x-1$ & $[ 3 3 3 1 ]$ & $[ 5 6 1 6 1 6 1 ]$ & \cite{src32}\\
  [.5ex]
  $[[60, 36, 2]]_{7^2}^{\circ}$ & $x^{60}-x-3$ & $[ 3 5 6 3 5 3 4 1 3 1 1 0 3 3 6 1 3 0 2 6 3 5 3 1 ]$ & $[ 2 1 ]$ & \cite{src13}\\
  [.5ex]
  $[[10, 2, 4]]_{13^2}^{\circ}$ & $x^{10}-2x-1$ & $[ 8 A 1 1 1 ]$ & $[ C 6 A 1 1 ]$ & \cite{src18}\\
  [.5ex]
    $[[15, 3, 5]]_{13^2}^{\circ}$ & $x^{15} - x^2 - 3$  & $[ 9 8 4 6 5 2 1 ]$  & $[ 5 9 6 0 C 5 1 ]$ & \cite{src18}\\
  [.5ex]
  $[[36, 28, 3]]_{13^2}^{\circ}$ & $x^{36}-x-4$ & $[ 5 0 2 1 ]$ & $[ 1 C 2 6 2 1 ]$ & \cite{src19}\\
  [.5ex]
  $[[36, 30, 2]]_{17^2}^{\circ}$ & $x^{36}-x-1$ & $[ 4 4 1 ]$ & $[ C A F 5 1]$ & \cite{src18}\\
  [.5ex]
\noalign{\smallskip}\hline
\end{longtable}
\hspace{-3.4cm}$^\circ$ delineates simpler construction codes.\\
%A-10, B-11, C-12, D-13, E-14, F-15, G-16

\end{document}